\documentclass[twocolumn,floatfix,prx,aps,showpacs]{revtex4-1}
\usepackage{graphicx,amsmath,amssymb,color}
\usepackage[utf8]{inputenc}
\usepackage{nicefrac}
\usepackage{multirow, makecell, ragged2e}
\usepackage[titletoc,title]{appendix}

\newcommand{\be}{\begin{equation}}
\newcommand{\ee}{\end{equation}}

\newcommand{\ba}{\begin{eqnarray}}
\newcommand{\ea}{\end{eqnarray}}

\renewcommand{\vec}[1]{\mbox{\boldmath$#1$}}

\begin{document}

\title{Exotic Bilayer Crystals in a Strong Magnetic Field}
\author{W. N. Faugno$^1$, A. J. Duthie$^2$, D. J. Wales$^2$, and J. K. Jain$^1$}

\affiliation{$^1$Department of Physics, 104 Davey Lab, Pennsylvania State University, University Park, Pennsylvania 16802, USA}
\affiliation{$^2$University Chemical Laboratories, Lensfield Road, Cambridge CB2 1EW, UK }

\date{\today}

\begin{abstract} 
Electron bilayers in a strong magnetic field exhibit insulating behavior for a wide range of interlayer separation $d$ for total Landau level fillings $\nu\leq 1/2$, which has been interpreted in terms of a pinned crystal. We study theoretically the competition between many strongly correlated liquid and crystal states and obtain the phase diagram as a function of quantum well width and $d$ for several filling factors of interest. We predict that three crystal structures can be realized: (i) At small $d$, the so-called triangular Ising antiferromagnetic (TIAF) crystal is stabilized in which the particles overall form a single-layer like triangular crystal while satisfying the condition that no nearest-neighbor triangle has all three particles in the same layer. (ii) At intermediate $d$, a correlated square (CS) crystal is stabilized, in which particles in each layer form a square lattice, with the particles in one layer located directly across the centers of the squares of the other. (iii) At large $d$, we find a bilayer graphene (BG) crystal in which the A and B sites of the graphene lattice lie in different layers. All crystals that we predict are strongly correlated crystals of composite fermions; a theory incorporating only electron Hartree-Fock crystals does not find any crystals besides the `trivial' ones occurring at large interlayer separations for total filling factor $\nu\leq1/3$ (when layers are uncorrelated and each layer is in the long familiar single-layer crystal phase). The TIAF, CS and BG crystals come in several varieties, with different flavors of composite fermions and different interlayer correlations. The appearance of these exotic crystal phases adds to the richness of the physics of electron bilayers in a strong magnetic field, and also provides insight into experimentally observed bilayer insulator as well as transitions within the insulating part of the phase diagram. 
\end{abstract}

\maketitle

\section{Introduction}

The rich physics of the fractional quantum Hall effect (FQHE) has been entangled with the search for a collective electron solid. For a two-dimensional electron gas (2DEG), a high magnetic field quenches the kinetic energy, suggesting that an electron crystal state ought to be realizable for filling factor $\nu<1$\cite{Wigner34, Lozovik75}. However, experiments reveal a liquid state, manifested through the FQHE\cite{Tsui82}. The FQHE has a rich phenomenology: A large number of fractions have been observed so far, most of which have the form $\nu=n/(2pn\pm 1)$ and $\nu=1-n/(2pn\pm 1)$. Calculations incorporating the physics of the FQHE predicted that the crystal should occur at filling factors $\nu<$1/6\cite{Lam84, Levesque84, Esfarjani90, Cote91, Zheng95, Goldoni96,  Yang01, Shibata03,  He05,  Shi07}.
Indeed, a large body of experimental work has shown a transition from the FQH liquid to an insulator at around $\nu=1/6$, with the insulating phase naturally interpreted as a crystal pinned by disorder
\cite{Goldman90, Jiang90, Jiang91, Williams91,Manoharan94a, Pan02, Engel97, Li00, Ye02, Chen04, Sambandamurthy06, Chen06}. Subsequent theoretical work clarified that the crystal is not an ordinary electron crystal, but rather a crystal of composite fermions, which provides an excellent representation of the crystal phase \cite{Yi98, Narevich01,Mandal03, Chang05, Archer13}. Recent experiments provide some evidence for the composite fermion (CF) nature of the crystal\cite{Liu14a,Zhang15c,Jang17}.

In this paper we study the nature of the crystal phase in bilayer systems. Bilayer systems can be realized either by fabricating two nearby quantum wells, or through a single wide quantum well (WQW) that behaves as a bilayer system for sufficiently large widths.  Previous theoretical investigations of bilayer states have focused primarily on the nature of two component liquid states, ignoring the electron crystal phases\cite{Chakraborty87, Yoshioka89, He91, He93, Scarola01b,Thiebaut14}. Many new FQH states become available as a function of the layer separation $d$. Such states have been considered in detailed theoretical calculations, and also studied experimentally. A striking example is the appearance of FQHE at total filling $\nu=1/2$\cite{Eisenstein92, Suen92b, Suen92,Suen94b}, which is understood in terms of the Halperin 331 state\cite{Halperin83}. (When used in the context of a bilayer system, $\nu$ will always refer to the {\em total} filling factor below.) Many phase transitions between various compressible and incompressible states have been predicted at each filling factor as a function of $d/l$, where $l=\sqrt{\hbar c/eB}$ is the magnetic length\cite{Narasimhan95, Scarola01b,Thiebaut14}. 

Multicomponent systems appear in many different contexts, where the components can be either the electron spin, relevant at low Zeeman energies, or the valley index in multivalley systems such as silicon, AlAs quantum wells, or graphene\cite{Shayegan06,Xu09,Apalkov10,Apalkov11}, or the layer index, as in bilayer systems. The bilayer systems in the limit of zero layer separation, when the interaction is independent of the layer index, are formally equivalent to the spin system at zero Zeeman energy. However, for nonzero layer separations the bilayer systems provide a way of tuning the inter-component interactions relative to the intra-component interactions, thus allowing realization of new physics not available to multi-component systems with SU(2) symmetry.

It can be expected that the crystal will also show a rich phase diagram in bilayer systems, with many competing liquid and crystal states appearing as a function of the interlayer separation and the filling factor. An interesting question is the nature of the crystal phase, and whether crystals other than a triangular crystal may be stabilized. This issue has been addressed theoretically in the past\cite{Narasimhan95, Goldoni96,Thiebaut15}, but without allowing for CF crystals\cite{Yi98,Narevich01,Chang05,Archer13}. For single layers, CF crystals are energetically more favorable than electron crystals, and necessary for explaining observed re-entrant phase transitions. An example is the theoretical explanation\cite{Archer13} of the re-entrant phase transitions observed \cite{Goldman90, Jiang90, Jiang91, Pan02} in the vicinity of $\nu=1/5$, where the system is insulating at $\nu<1/5$ and for a range of $\nu$ between 1/5 and 2/9, but exhibits FQHE at $\nu=1/5$ and $\nu=2/9$. As we shall see below, allowing for CF crystals will be crucial for identifying bilayer crystal states.

The primary motivation for our study comes from experiments. In their study of bilayer systems, Eisenstein {\em et al.} \cite{Eisenstein92} found that the system becomes insulating in the vicinity of total filling $\nu=1/2$, although it exhibits a FQH state at $\nu=1/2$ for small interlayer separations. 
Magnetotransport experiments in WQWs carried out by 
Manoharan {\em et al.}\cite{Manoharan96} explored a large region of parameter space in terms of the two-dimensional electron density and filling factors. They also found that the insulating phase dominates for a large range of parameters for total filling $\nu\leq 1/2$. Shabani {\em et al.}\cite{Shabani13} have performed an extensive study of the phase diagram at $\nu=1/2$ in WQWs. Microwave spectroscopy has also been used to characterize the insulating states in the WQW systems\cite{Chen04, Hatke15, Hatke14, Wang17b}, to reveal structure that is inaccessible in DC magnetotransport experiments. Sharp resonances are seen for the insulating phases, which are interpreted as pinning modes of a crystal. One of the interesting findings has been shifts in the resonant frequency inside the insulating region of the phase diagram, which the authors have taken as evidence that there may be a reordering of the crystal configuration\cite{Hatke15, Hatke14, Wang17b}. It is therefore of interest to identify what kinds of crystals are feasible in bilayer systems.

\begin{figure}
\includegraphics[width = 3.7in]{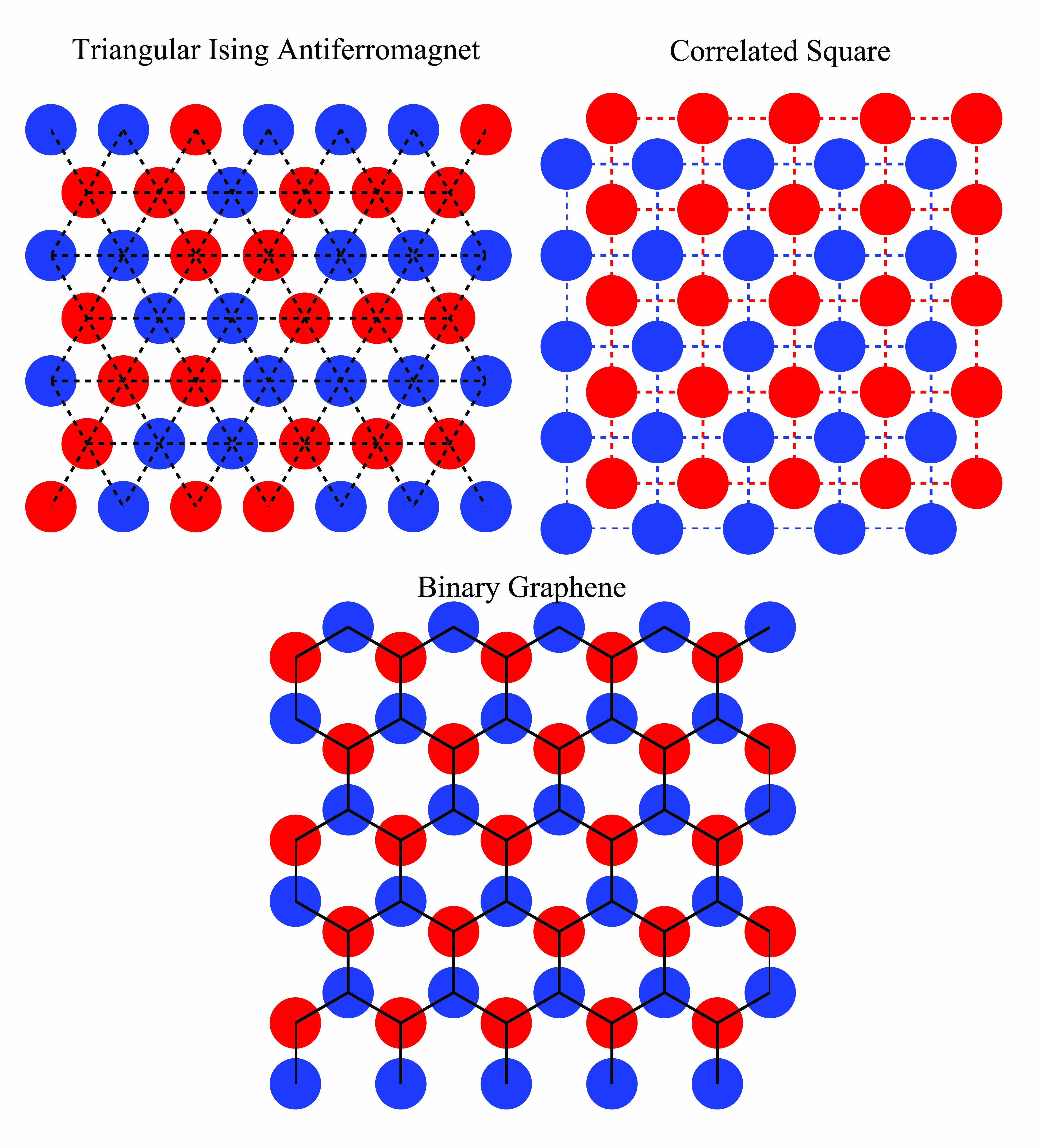}
\caption{Two-dimensional lattices considered in our work. The blue and red colors denote different layer indices. Triangular Ising AntiFerromagnetic (TIAF) crystal is a triangular lattice with half of the particles in one layer and half in the other, such that each triangle contains two particles in one layer and one in the other. In the Correlated Square (CS) lattice each layer forms a square lattice whose sites are aligned with the centers of the squares in the opposite layer. Finally, the Binary Graphene (BG) crystal has the overall structure of graphene, but with the A and B sublattices lying in different layers. We have chosen these configurations because they are the lowest energy solutions to the classical bilayer Thomson problem for different ranges of layer separation. }
\label{2Dreps}
\end{figure}

We consider in this work electron and composite fermion crystals (CFCs) in addition to the FQH liquid states. We determine the energies of a large class of variational wave functions for the liquid and crystal phases to determine the lowest energy state as a function of the layer separation $d/l$. We predict three new crystal phases in bilayer systems, shown in Fig.~\ref{2Dreps}:

\begin{itemize}

\item Triangular Ising antiferromagnetic (TIAF) crystal: When viewed from above, this looks like a single layer triangular crystal, but half of the particles are in one layer and half in the other satisfying the condition that no nearest-neighbor triangle has all three particles in the same layer.

\item Correlated square (CS) crystal: This crystal consists of two interpenetrating square lattices, such that the sites in one layer lie across the centers of the squares in the opposite layer. 

\item Binary graphene (BG) crystal: This crystal, when viewed from above, looks like a graphene lattice, with the A and B lattice sites residing in different layers.

\end{itemize}

The CS and BG crystals were also considered previously by Thiebaut, Regnault and Goerbig \cite{Thiebaut15} in their Hartree-Fock study of the crystal phase at $\nu=1/2$ in the lowest and the first excited Landau levels (LLs) in WQWs.

\begin{figure*}
\includegraphics[width = 7in]{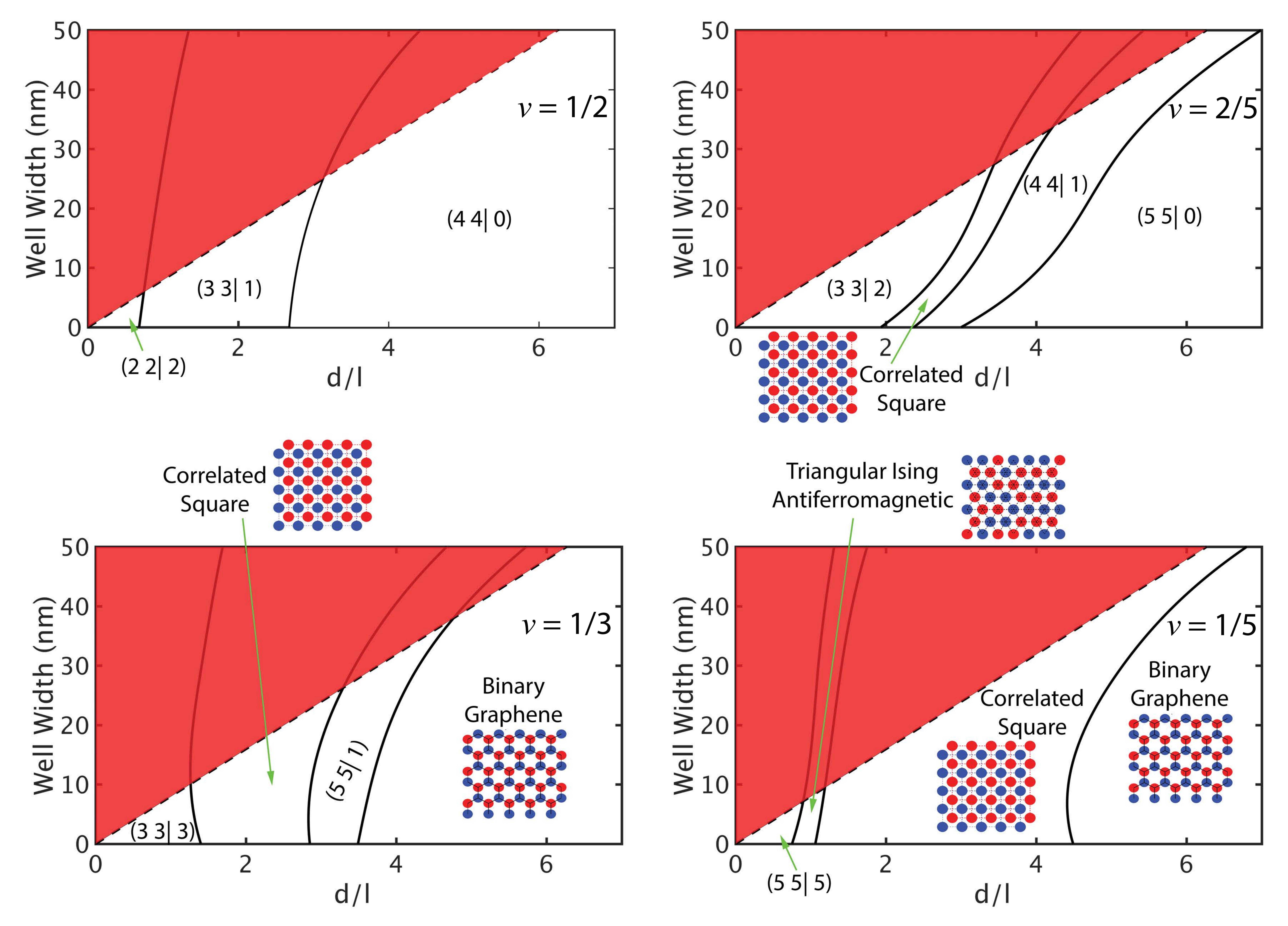}
\caption{Phase diagram of liquid and crystal phases as a function of the quantum well width and the interlayer separation $d/l$. 
To avoid clutter, we have suppressed states that occur in very narrow ranges of parameters, and we have also omitted the nature of interlayer correlations in this figure. These finer details can be found in what follows. This phase diagram corresponds to the density of $10^{11}$cm$^{-2}$, and assumes parameters appropriate for GaAs quantum wells. The shaded region above the dashed line is unphysical since here the quantum well width exceeds the interlayer separation.}
\label{summary}
\end{figure*}

Before we come to the calculational details, we show in Fig.~\ref{summary} the phase diagrams for several total filling factors as a function of the quantum well width and $d/l$ for a system with electron density of 10$^{11}$ cm$^{-2}$. This representation captures the general behavior found for other parameters, although the details of the phase boundary vary. (Many fine details regarding the correlations of the crystals have been suppressed here for simplicity; they are given later in the article.) The appearance of the three crystal states as a function of $d/l$ can be understood intuitively.  For small $d/l$, the inter and intralayer interactions are approximately equal. A triangular crystal forms as though the system were a single layer, and the two layers are accommodated through a frustrated ``pseudospin" structure. At intermediate separations, when the intra-layer correlations become relatively weak, the CS crystal appears, which builds good interlayer correlations between particles, as also found in Hartree-Fock studies \cite{Narasimhan95, Zheng95}.  Finally, for large separations, the layers act almost independently and form two triangular crystals within their respective layers, but the weak interlayer interaction stabilizes the BG lattice. Results for filling factors at several densities are presented in detail later in section V. 

We stress that the TIAF, CS and BG crystals can each come in several varieties, with different flavors of composite fermions and different interlayer correlations. Their full identification will require two other integers (which have been suppressed in Fig.~\ref{summary} to avoid clutter). 

It should be stressed that {\em all of the bilayer crystals we find are {\em CF} crystals}. No crystals would be stabilized if we only worked with electron Hartree-Fock crystals, with the trivial exception of the large $d$ phase at total filling $\nu\leq 1/3$ where electrons in each individual layer have filling factor $\leq 1/6$ and thus form the  long familiar single layer crystal. The CF physics is thus crucial for stabilizing crystals with inherently bilayer character.

The paper is structured as follows. In section II, we present a general background for the theory used to construct the wave functions. We then describe the method for obtaining the crystal coordinates in a spherical geometry in section III. Section IV outlines our computational method. In section V, we present results for a quantitative study of FQH systems in a bilayer, focusing on zero width and double quantum well systems. In section VI, we conclude by comparing with existing experiments, and make predictions for future experiments.

\section{Model states}

\begin{table}
\begin{center}
\begin{tabular} { | c | c | c | }
\hline
\multicolumn{3}{|c|}{Liquid states and wave functions}\\
\hline
$\nu$ & State & wave function\\
\hline
\multirow{3}{2em}{\LARGE $\frac{1}{2}$} & $(4\ 4|\ 0)$ & $\Psi_{\frac{1}{4}}(\{z_i\})\Psi_{\frac{1}{4}}(\{w_i\})$\\
& $(3\ 3|\ 1)$ & $\Psi_{\frac{1}{3}}(\{z_i\})\Psi_{\frac{1}{3}}(\{w_i\})\Pi_{i,j}(z_i-w_j)$\\
& $(2\ 2|\ 2)$ & $\Psi_{\frac{1}{2}}(\{z_i\})\Psi_{\frac{1}{2}}(\{w_i\})\Pi_{i,j}(z_i-w_j)^2$\\
\hline
\multirow{3}{2em}{\LARGE $\frac{2}{5}$} & $(5\ 5|\ 0)$ & $\Psi_{\frac{1}{5}}(\{z_i\})\Psi_{\frac{1}{5}}(\{w_i\})$\\
& $(4\ 4|\ 1)$ & $\Psi_{\frac{1}{4}}(\{z_i\})\Psi_{\frac{1}{4}}(\{w_i\})\Pi_{i,j}(z_i-w_j)$\\
& $(3\ 3|\ 2)$ & $\Psi_{\frac{1}{3}}(\{z_i\})\Psi_{\frac{1}{3}}(\{w_i\})\Pi_{i,j}(z_i-w_j)^2$\\
\hline
\multirow{4}{2em}{\LARGE $\frac{1}{3}$} & $(6\ 6|\ 0)$ & $\Psi_{\frac{1}{6}}(\{z_i\})\Psi_{\frac{1}{6}}(\{w_i\})$\\
& $(5\ 5|\ 1)$ & $\Psi_{\frac{1}{5}}(\{z_i\})\Psi_{\frac{1}{5}}(\{w_i\})\Pi_{i,j}(z_i-w_j)$\\
& $(4\ 4|\ 2)$ & $\Psi_{\frac{1}{4}}(\{z_i\})\Psi_{\frac{1}{4}}(\{w_i\})\Pi_{i,j}(z_i-w_j)^2$\\
& $(3\ 3|\ 3)$ & $\Psi_{\frac{1}{3}}(\{z_i\})\Psi_{\frac{1}{3}}(\{w_i\})\Pi_{i,j}(z_i-w_j)^3$\\
\hline
\multirow{4}{2em}{\LARGE $\frac{1}{5}$} & $(10\ 10|\ 0)$ & $\Psi_{\frac{1}{10}}(\{z_i\})\Psi_{\frac{1}{10}}(\{w_i\})$\\
& $(9\ 9|\ 1)$ & $\Psi_{\frac{1}{9}}(\{z_i\})\Psi_{\frac{1}{9}}(\{w_i\})\Pi_{i,j}(z_i-w_j)$\\
& $(8\ 8|\ 2)$ & $\Psi_{\frac{1}{8}}(\{z_i\})\Psi_{\frac{1}{8}}(\{w_i\})\Pi_{i,j}(z_i-w_j)^2$\\
& $(7\ 7|\ 3)$ & $\Psi_{\frac{1}{7}}(\{z_i\})\Psi_{\frac{1}{7}}(\{w_i\})\Pi_{i,j}(z_i-w_j)^3$\\
& $(6\ 6|\ 4)$ & $\Psi_{\frac{1}{6}}(\{z_i\})\Psi_{\frac{1}{6}}(\{w_i\})\Pi_{i,j}(z_i-w_j)^4$\\
& $(5\ 5|\ 5)$ & $\Psi_{\frac{1}{5}}(\{z_i\})\Psi_{\frac{1}{5}}(\{w_i\})\Pi_{i,j}(z_i-w_j)^5$\\
\hline
\end{tabular}
\end{center}
\caption{CF liquid wave functions. This table lists all filling factors and liquid states considered in this study. The wave function $\Psi_{\bar\nu}(\{z_i\})$ at $\bar\nu=\frac{n}{2pn+1}$ is defined in the text, and $m$ is the number of interlayer zeros. Wave functions are labeled $(\bar\nu^{-1}\ \bar\nu^{-1}|\ m)$.}
\end{table}

\begin{table*}
\begin{center}
\begin{tabular} {|c|c|c|}
\hline
\multicolumn{3}{|c|}{Crystal notation and wave functions}\\
\hline
Notation & Name & wave function\\
\hline
BG$(2p,m)$ & \makecell{Binary Graphene\\CF crystal} & $\Psi^{\text{BG}(2p)}_{\bar\nu}(\{z_i\})\Psi^{\text{BG}(2p)}_{\bar\nu}(\{w_i\})\Pi_{i,j}(z_i - w_j)^m$\\
\hline
CS$(2p,m)$ & \makecell{Correlated Square\\CF crystal} & $\Psi^{\text{CS}(2p)}_{\bar\nu}(\{z_i\})\Psi^{\text{CS}(2p)}_{\bar\nu}(\{w_i\})\Pi_{i,j}(z_i-w_j)^m$\\
\hline
TIAF$(2p,m)$ & \makecell{Triangular Ising\\Antiferromagnetic\\CF crystal} & $\Psi^{\text{TIAF}(2p)}_{\bar\nu}(\{z_i\})\Psi^{\text{TIAF}(2p)}_{\bar\nu}(\{w_i\})\Pi_{i,j}(z_i-w_j)^m$\\
\hline
\end{tabular}
\end{center}
\caption{This table lists the form for all crystal wave functions considered in the article. 
$\Psi^{X(2p)}_{\bar\nu}$ is the wave function of the LLL crystal of $^{2p}$CFs at filling $\bar{\nu}$, and the integer $m$ represents the strength of the interlayer correlations. Superscripts BG, CS, and TIAF correspond to Binary Graphene, Correlated Square and Triangular Ising Antiferromagnet. The representations of these crystals on a bilayer sphere are obtained through analysis of the bilayer Thomson problem. 
}
\end{table*}

For our study, we will consider several liquid and crystal wave functions from CF theory. These wave functions have been demonstrated to be very accurate in describing the physics, in single layers, of both liquids\cite{Jain89,Jain07} and crystals\cite{Chang05}. We begin each section by describing the construction of the single layer wave functions, followed by bilayer wave functions. Unlike the single layer crystals where the triangular lattice is the only (known) energetically favorable configuration, multiple lattice structures can be realized in bilayer systems, depending on the layer separation and the filling factor.

\subsection{CF theory of the FQH liquid}

Composite fermions are bound states of electrons and an even number ($2p$) of vortices\cite{Jain89,Jain07,Lopez91,Halperin93}. Composite fermions are weakly interacting, and experience an effective magnetic field $B^*=B-2p\rho\phi_0$, where $\phi_0=hc/e$ is a flux quantum and $\rho$ is the 2D electron or CF density.  They form LL-like levels referred to as $\Lambda$ levels ($\Lambda$Ls), and fill $\nu^*$ of them, where $\nu=\nu^*/(2p\nu^*\pm 1)$. The FQHE at $\nu=n/(2pn\pm 1)$ is a manifestation of the integer quantum Hall effect (IQHE) of weakly interacting composite fermions at CF filling $\nu^*=n$. The composite fermions with $2p$ vortices bound to them are denoted as $^{2p}$CFs.

For fully spin polarized electrons in a single layer, the Jain CF wave function for the ground state at $\nu=n/(2pn+1)$ is given by 
\begin{equation}
\Psi_{\frac{n}{2np+1}} = P_{\rm LLL}\Phi_n \Pi_{i<j}(z_i - z_j)^{2p}
\end{equation}
where $\Phi_n$ is the wave function for electrons at $\nu^*=n$ and $z_i = x_i -iy_i$ are the coordinates of the $i$th electron. 
$P_{\rm LLL}$ denotes lowest Landau level (LLL) projection, which will be evaluated numerically via the Jain-Kamilla method\cite{Jain97}. For the ground state at $\nu=1/(2p+1)$, i.e. for $\nu^*=1$, this wave function reproduces the Laughlin wave function.

The above construction can be generalized straightforwardly to a system of spinful electrons in a single layer\cite{Wu93, Park98, Mandal96}. Here we have $n = n_{\uparrow}+n_{\downarrow}$,  where $n_{\uparrow}$ and $n_{\downarrow}$ are the numbers of occupied spin up and spin down  $\Lambda$ levels. 
Since the interaction is spin independent, the ground state wave function is an eigenstate of the total spin operator $\bold{S}^2 = (\Sigma_i^{N_{tot}}\bold{S}_i)^2$, where $\bold{S}_i$ is the spin operator acting on the $i$th particle and $N_{tot}$ is the total number of particles. The Jain wave functions for spinful composite fermions at $\nu=n/(2pn+1)$ are given by
\begin{equation}
\Psi_{\frac{n}{2np+1}}^{(n_\uparrow,n_\downarrow)} = A[P_{\rm LLL}\Pi_{i<j}(z_i - z_j)^{2p}\Phi_{n_{\uparrow}}\Phi_{n_{\downarrow}}\alpha_1...\alpha_{N_1}\beta_1...\beta_{N_2}]
\end{equation}
where $A$ is the antisymmetrization operator, $N_1$ and $N_2$ are the numbers of composite fermions with up and down spins, and $\alpha$ and $\beta$ are up and down spinors. This wave function satisfies the Fock cyclic conditions with total spin quantum number $S=S_z=(N_1-N_2)/2$\cite{Girvin07}. 
Spinful electrons in general have several states at any given filling factor due to the freedom to choose different combinations of $n_\uparrow$ and $n_\downarrow$. At zero Zeeman energy, the ground state corresponds to $n_\uparrow=n_\downarrow=n/2$ for even $n$, while for odd $n$ we have $n_\uparrow = (n+1)/2$ and $n_\downarrow = (n-1)/2$. In the special case of $n=1$, a fully spin polarized state is obtained with $n_\uparrow=1$ and $n_\downarrow=0$. 
 
We now come to bilayer systems.  A bilayer system with zero layer separation ($d/l=0$) is formally equivalent to the spin degree of freedom in a single layer system with Zeeman energy set to zero\cite{Jain89, Jain07}, with the two layers representing spin up and spin down. This follows because the interaction is independent of the layer index in this limit, so the Hamiltonian satisfies the exact SU(2) symmetry. The bilayer degree of freedom is sometimes referred to as the pseudospin.

The layer pseudospin degree of freedom can create further new structures for $d/l\neq 0$ because the interaction becomes pseudospin dependent, and the wave function no longer needs to satisfy the Fock conditions. Following Scarola and Jain\cite{Scarola01b}, we consider here the following class of wave functions
\begin{equation}
\Psi_\nu^{(\bar\nu^{-1} \bar\nu^{-1} |m)} = \Pi_{i,j}(z_i - w_j)^m\Psi_{\bar\nu}(\{z_i\}) \Psi_{\bar\nu}(\{w_i\})
\label{bilayer_wf}
\end{equation}
where $\{z_i\}$ and $\{w_i\}$ are the coordinates of particles in different layers, and we have assumed equal carrier densities in each layer.  We take for the single layer wave function $\Psi_{\bar\nu}(\{z_i\})=P_{\rm LLL}\prod_{j<k}(z_j-z_k)^{2p}\Phi_{n}$ with $\bar{\nu}=n/(2pn+1)$. The factor $\Pi_{i,j}(z_i - w_j)^m$ introduces correlation between the layers through interlayer vortices. The total filling factor $\nu$ is given by\cite{Scarola01b}
\begin{equation}
\nu = \frac{2\bar\nu}{1+m\bar\nu}
\end{equation}
We can now enumerate all the candidate states for a given total filling factor. We consider $m\leq 2p+1$ because $m>2p+1$ would represent stronger interlayer correlations than intralayer correlations, which is physically unreasonable. The limiting form for $d/l=0$ is known from the spin problem described previously.

In this article we will consider total filling factors $\nu=1/2$, 2/5, 1/3, and 1/5. Table 1 enumerates all of the liquid states of the form given in Eq.~\ref{bilayer_wf} at these filling factors. For $\bar\nu=1/(2p+1)$ these wave function reduce to the Halperin wave functions\cite{Halperin83}.

The above wave functions are written for the planar geometry. 
For our calculations, we work in the spherical geometry to avoid potential problems resulting from edge effects on disks\cite{Yi98, Haldane83}.
We confine our particles to a spherical shell with a magnetic monopole of strength $Q$ placed at the center generating a radial magnetic field. The value of $2Q$ is restricted to be an integer, equal to the number of flux quanta penetrating the surface of the sphere. The radius of the sphere is $2\sqrt{Q}l$.  
When considering the FQHE in spherical geometry, we follow Haldane \cite{Haldane83} to define spinor coordinates $u_i$ and $v_i$
\begin{equation}
\begin{gathered}
u_i=\cos(\theta_i/2)e^{i\phi_i/2}\\
v_i=\sin(\theta_i/2)e^{-i\phi_i/2}
\end{gathered}
\end{equation}
where $\theta$ and $\phi$ are the angular coordinates.
The wave function is then written as
\begin{equation}
\Psi_\nu^{(\bar\nu^{-1}\bar\nu^{-1}|m)}=\Pi_{i,j}(u_iv_j-u_jv_i)^m\Psi_{\bar\nu}(\{z_i\})\Psi_{\bar\nu}(\{w_i\})
\end{equation}
The single particle states in $\Psi_{\bar\nu}$ are the monopole harmonics $Y_{Q^*,l,m}$ where $Q^*$ is the effective magnetic monopole strength and $l = |Q^*| + n$ with $n$ the number of the current $\Lambda$L. The index $m$ is restricted to be between $\pm l$\cite{Jain97}. 
The above bilayer wave functions correspond to the total flux
\cite{Scarola01b}
\begin{equation}
2Q = \frac{(2pn + mn +1)N - (2pn+n^2)}{n}
\end{equation}
We assume here and below the notation in which the total number of particles in a bilayer is $N_{\rm tot}=2N$, so that each layer individually has $N$ particles. 

\subsection{CF crystal states}

We begin with the CF crystal (CFC) wave function for a single layer system. Because it is not possible to fit a triangular crystal perfectly on the surface of a sphere, we consider a ``Thomson crystal," where the lattice positions are determined by finding the lowest energy configuration of 
classical point charges on a sphere. 
More details on the Thomson problem are given in the following section. We denote the Thomson crystal positions as 
\be
(U_i, V_i)= (\cos(\gamma_i/2)e^{i\delta_i/2}, \sin(\gamma_i/2)e^{-i\delta_i/2})
\ee
In a spherical geometry, the wave function for a Gaussian wave packet localized at $(U, V)$ is given by $(U^* u + V^* v)^{2Q^{*}}$ for a system at flux $2Q^*$.
The CFC wave function is then given by\cite{Archer13}
\be
\Psi^{X(2p)}_\nu(\{u_i, v_i\}) = \det(U_i^* u_j + V_i^* v_j)^{2Q^{*}}\Pi_{i<j}(u_iv_j-u_jv_i)^{2p}
\ee
where $U_i$ and $V_i$ are the spinors corresponding to each lattice site at coordinates $(\gamma_i,\delta_i)$. These wave functions are by construction in the LLL. The symbol $X(2p)$ denotes different possible crystal structures of composite fermions carrying $2p$ vortices.

We now form bilayer crystal wave functions:
\begin{multline}
\Psi^{X(2p,m)}_\nu=\Psi^{X(2p)}_{\bar{\nu}}(\{u_{1,i}, v_{1,i}\}) \Psi^{X(2p)}_{\bar{\nu}}(\{u_{2,i}, v_{2,i}\})\\
 \Pi_{i,j}(u_{1,i}v_{2,j} - u_{2,j}v_{1,i})^{m}
\end{multline}
In this notation, $X(2p,m)$ refers to a bilayer crystal of type $X$ (which can be ``TIAF," ``CS" or ``BG") of composite fermions carrying $2p$ vortices, with $m$ interlayer zeros. The filling factor $\bar\nu$ is given by $\bar\nu=N/(2Q^*+2p(N-1))$.
The positions of the crystal lattice sites are determined by solving the bilayer Thomson problem (see next section for further details).

We will determine the lowest energy state out of all candidate states as a function of various parameters. 
For bilayer systems, we consider the effective interaction
\begin{equation}
V_{\uparrow\uparrow}(\vec{r_i},\vec{r_j}) = V_{\downarrow\downarrow}(\vec{r_i},\vec{r_j}) = \frac{1}{|\vec{r_i} - \vec{r_j|}}
\label{interaction1}
\end{equation}
\begin{equation}
V_{\uparrow\downarrow}(\vec{r_i},\vec{r_j}) = \frac{1}{\sqrt{|\vec{r_i} - \vec{r_j}|^2 + d^2}}
\label{interaction2}
\end{equation}
where $d$ is the distance between the layers  and the arrows label the pseudospin corresponding to left and right layers. We denote all lengths in units of the magnetic length $l$ and energies in units of $e^2/\epsilon l$. We have assumed that there is no nearby conducting layer to screen the Coulomb interaction within our bilayer system. 

For a proper comparison, the crystal state must correspond to the same filling factor as the liquid state. We accomplish this by using the same number of particles as well as the same value for the physical magnetic flux $2Q$. We construct multiple states at filling factor $\nu$ by considering all values of $2p$ and $m$ such that $2Q^* = 2Q - 2p(N-1) - mN$ is nonnegative and $2p \geq m$. For a full summary of the states we have studied, see Tables 1 and 2. We stress that we confine our search to the crystal structures that appear prominently in the bilayer Thomson problem. 

\section{Thomson Crystal for a Bilayer System}

A crucial task is to determine what are the most promising crystal configurations for the bilayer problem, and also the best representations of these crystals on a sphere. For this, a variant of the classical Thomson problem to include two types of charged particles was studied.
The resulting low-energy configurations, created in the absence of magnetic fields, were then used as seeds for the more detailed magnetic field calculations.

Finding the lowest energy arrangement of $N$ classical point charges confined to the surface of a sphere is known as the Thomson problem\cite{thomsonproblem}. For $N$ = 2--6 and 12, analytical solutions are known. These values are significant, as the structures are invariant if the Coulombic potential is replaced with a limiting potential of the form $V(\tilde{r})=\lim_{n\to\infty}\tilde{r}^{-n}$, or a logarithmic interaction\cite{eqmconfigsErber}, where $\tilde{r}$ is the distance between the charged particles. Solving the problem with the first of these potentials corresponds to the Tammes problem\cite{tammesproblem} of packing $N$ particles on the surface of a sphere whilst maximising all particle-particle arc lengths. This potential invariance reveals the power of symmetry as a structural determinant for small $N$, though computational methods must be used for larger $N$ as the geometrical symmetry is lost\cite{eqmconfigsErber}.

In previous work, the Thomson problem has been used as an approximate basis for designing carbon cages larger than the stable truncated icosahedron form of C$_{60}$. 860 and 1160 particle Thomson problem minima were used as starting points for C$_{860}$ and C$_{1160}$, and minimized using density functional theory\cite{Bowick,defectmotifs}. This study highlights the utility of the Thomson problem minima as starting points for more detailed calculations.

The process of finding energy minima for different systems employs geometry optimisation. For a given configuration of particles and an arbitrary potential between them, local optimisation produces a minimum on the potential energy surface (PES). The global minimum is the minimum with the lowest energy. Even small systems, such as a cluster of 38 Lennard-Jones atoms\cite{walesbook}, have a large number of minima\cite{lj38,nomin}, and enumerating all of them is usually either not possible or an extremely inefficient way of locating the global minimum.

Global optimisation for Thomson systems is complicated by the fact that there are many metastable states separated by only small energy differences, with the number of minima rising exponentially with $N$\cite{eqmconfigsErber,Bowick,defectmotifsmeancurvature}. However, basin-hopping global optimisation \cite{Li6611,basinhopping} has been effective for selected $N$ up to 4352\cite{Bowick,defectmotifs}. In this approach, steps are taken between local minima, and are accepted or rejected based on a Metropolis condition with a fictitious temperature parameter. 

Perfect 2D \emph{hexagonal} close-packed structures cannot be bent to exist on the surface of a sphere, and so defects must be introduced in the Thomson problem minima. It is not possible to transform a 2D surface into a spherical form without cuts or distortions, which here manifest as different coordination sites. If the number of nearest-neighbours of a particle is C, then a disclination charge, Q, can be defined as Q = 6 - C. Euler's theorem\cite{eulerstheorem} states that the total disclination charge must be equal to 12 for close-packed structures on the surface of a sphere. There are many ways in which Euler's theorem can be satisfied, and the Thomson problem has been studied for thousands of particcles\cite{Bowick,defectmotifs}. The presence and nature of these defect motifs is central to determining system properties in the presence of external forces, and can aid understanding of macroscopic systems\cite{sphericalcrystals}.

Here, the binary, or bilayer, Thomson problem is considered, in which two types of charged particles are confined to the surface of a sphere. The interactions within each group are Coulombic, but the interactions between particles in different groups have a damped form, with the damping strength determined by an adjustable parameter, $\delta$, the interlayer separation. We note that this $\delta$ is not the same as $d/l$. The pairwise potential for $N$ particles on a sphere of radius $R$ is:

\begin{equation}
V_{ij}^{bin} = \begin{cases} \displaystyle\frac{1}{\displaystyle(\frac{\tilde{r}_{ij}}{R})}, & \text{for }  i,j  \text{ in the same group.}\\  \displaystyle\frac{1}{\sqrt{(\displaystyle\frac{\tilde{r}_{ij}}{R})^2 + (\frac{\delta}{N})^2}},& \text{for } i,j  \text{ in different groups.}
\end{cases} \label{potentialdefinition}
\end{equation}
In the potential, $\tilde{r}_{ij}/R$ is used as $\tilde{r}_{ij}$ is measured in units of the sphere radius $R$. The ratio $\delta/N$ can be considered as the separation between two infinite bilayers, which is the limit for a sphere of infinite radius. The adjustable parameter $\delta$ is scaled by $N$ as behaviour is expected to change on a length scale comparable to the interparticle separation. The aim of this scaling was to align similar regimes of behaviour to similar values of $\delta$ for different system sizes.

Following the success of basin-hopping global optimisation for the regular Thomson problem\cite{ssp}, the same approach was used here to locate the global minima for a variety of different compositions. The GMIN program\cite{software} was employed for the basin-hopping calculations, using the L-BFGS (Limited-memory BFGS) algorithm\cite{L-BFGS} for  energy minimisation. The energies of the minima are not changed by the basin-hopping algorithm, but downhill transition state barriers are removed, which allows more rapid sampling of the energy landscape. The use of the basin-hopping algorithm in combination with combinatorial searching\cite{combsearch} allows for efficient relaxation to the global minimum in multicomponent systems\cite{globopt}.

 For 45 particles of each type, around 50,000 basin-hopping steps were required to achieve convergence to the same minimum from 10 random starting points. The number of steps required decreases as the systems are made smaller, since there are fewer minima on the energy landscape. The proposed global minima for different compositions were used as seeds for the calculations in section IV. Tuning the interlayer separation provided three sets of coordinates to consider, corresponding to the BG, CS, and TIAF crystals.

\section{Technical details}

We determine the best variational ground state for the pseudospin dependent interaction in Eqs. \ref{interaction1} and \ref{interaction2} by calculating the energies for a series of trial wave functions of the of the form presented in Tables 1 and 2. 
We compute the energy expectation value, which is a $4N$ dimensional integral (recall we have $N_{\rm tot}=2N$ particles), by the Monte Carlo method, which allows us to determine the 
energy with up to $0.01\%$ accuracy with $10^7$ iterations. Using this method, we have calculated energies for total particle numbers up to $2N=98$. We calculate the energy for several system sizes and use a linear extrapolation to obtain the thermodynamic energy for every candidate state. The errors quoted below originate primarily from the uncertainty in the linear fit; the Monte Carlo simulation error for each energy is typically smaller by an order of magnitude. The fitting error is particularly significant for crystals as they necessarily have some defects due to the curvature. 

To obtain an energy value that is intensive, it is necessary to consider the total energy, including the background-background and electron-background interactions. In our case, since we are interested in comparing states, we measure the electron-electron Coulomb interaction relative to one of the candidate states.

Some of our wave functions will involve compressible composite fermion Fermi sea, for which we will use total particle numbers $2N=$ 18, 32, 50, 72, and 98, so that the effective magnetic field vanishes in each layer.

The total filling factor in spherical coordinates is defined to be $\nu=\lim_{N\rightarrow \infty}\frac{2N}{2Q}$ where $N$ is the number of particles in a single layer. 
Due to the finite size shift in the spherical geometry, the density for a finite $N$ is not the same as that in the thermodynamic limit, which provides an $N$ dependent correction to the energy. To 
compensate for this effect
we multiply the energy by the ratio of the interparticle separation in the thermodynamic limit to that in the finite system, i.e. $\sqrt{\frac{\rho_\infty}{\rho_N}} = \sqrt{\frac{2Q\nu}{2N}}$. This density correction reduces the dependence of the energy on the particle number, thus facilitating the comparison between the different candidate states\cite{Jain07}.

To connect with experimental systems, we also consider 2DEGs with finite width. 
We consider a double quantum well geometry, consisting of two wells of equal width. The effective intra-layer and interlayer Coulomb interactions are of the form
\begin{equation}
V_{\uparrow,\uparrow \ \rm eff} (r)= \frac{e^2}{\epsilon} \int d\zeta_1 \int d\zeta_2 \frac{|\xi(\zeta_1)|^2 |\xi(\zeta_2)|^2}{\sqrt{r^2 + (\zeta_1 - \zeta_2)^2}}
\end{equation}
\begin{equation}
V_{\uparrow,\downarrow \ \rm eff} (r)= \frac{e^2}{\epsilon} \int d\zeta_1 \int d\zeta_2 \frac{|\xi(\zeta_1)|^2 |\xi(\zeta_2)|^2}{\sqrt{r^2 + (\zeta_1 - \zeta_2 + d)^2}}
\end{equation}
where $\zeta_i$ is the distance perpendicular to the 2DEG and $\vec{r}$ is the coordinate in the plane of the 2DEG. The transverse component of the wave function, $\xi$, is obtained via self-consistently solving the Schr\"odinger and Poisson equations and applying the local density approximation (LDA). To carry out these calculations, we only need to know the shape of the confinement potential and the density of electrons. We have calculated the energies in the zero width limit and for double quantum well widths, 180\AA, 300\AA, 400\AA, and 500\AA. For further details on how the finite width calculation is carried out we refer the reader to Ref.~\cite{Park99b}.

\section{Results}

We now present our results for total filling factors $\nu=$1/3, 2/5, 1/2 and 1/5. As defined in section II, our notation is $(\bar\nu^{-1}\ \bar\nu^{-1}|\ m)$ for liquid states, and $X(2p, m)$ for crystal states. $X=$ CS, BG, and TIAF represent correlated square, binary graphene, and triangular Ising antiferromagnetic lattices, respectively. The integers $2p$ and $m$ correspond to the CF vorticity and the number of interlayer correlation zeros.

\subsection{Zero Width}

\begin{figure*}
\includegraphics[width = \linewidth]{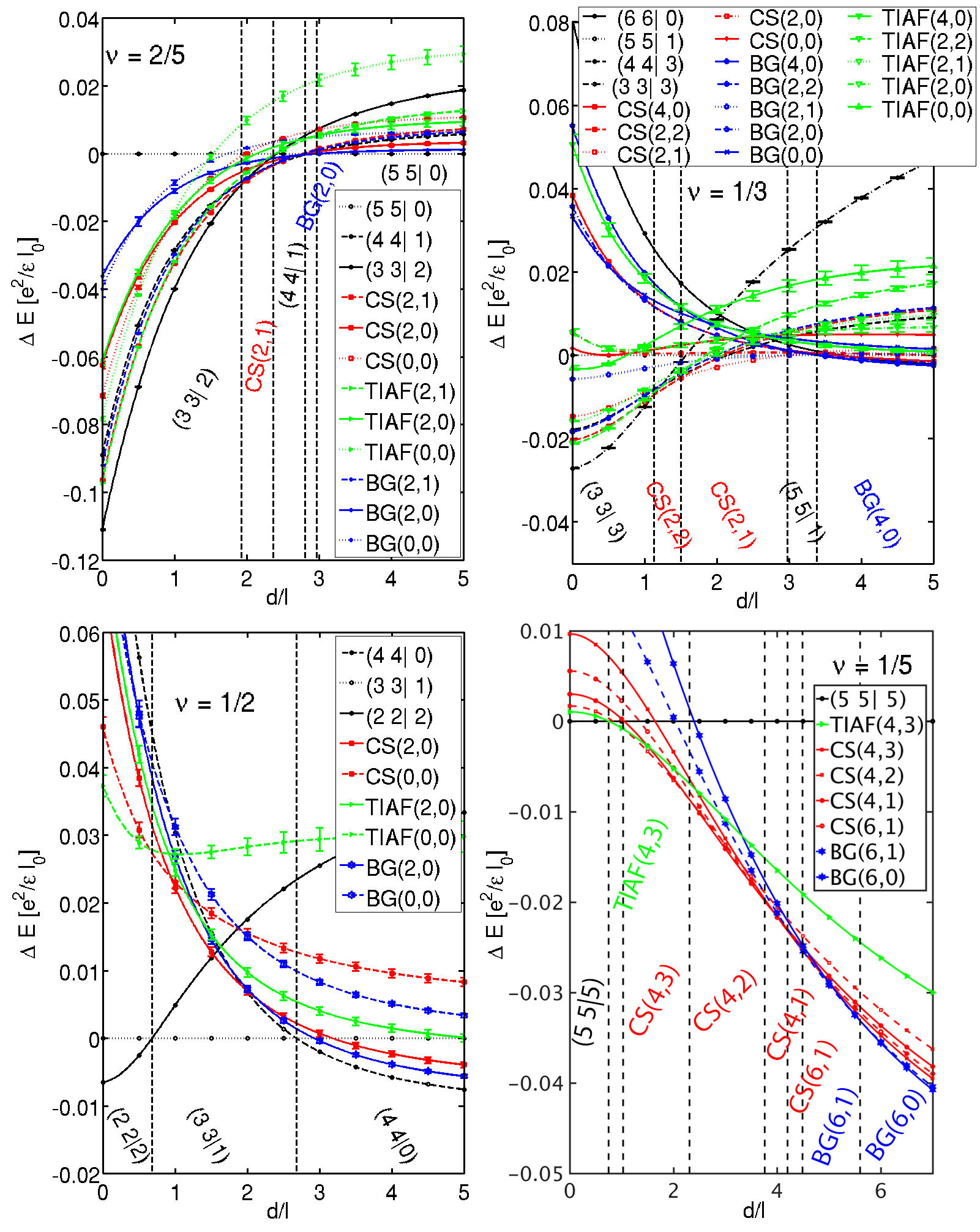}
\caption{Energies of bilayer liquid and crystal states for zero width layers as a function of the interlayer separation. The energy of each state is measured relative to a chosen reference incompressible liquid state (which itself appears as the zero energy state). All energy differences represent the thermodynamic limits, obtained as described in section IV.  The vertical dashed lines separate different ground state phases labeled on the plot. Here black corresponds to liquid states, while red, blue and green denote CS, BG and TIAF crystals.}
\label{0width}
\end{figure*}

We first consider a bilayer system with each layer of zero width. Figure \ref{0width} shows energies of various states at $\nu=1/3$, 2/5, 1/2 and 1/5 as a function of layer separation. At each filling, the energies are quoted relative to the energy of a reference state, which itself shows up as the zero energy state in our plots. Level crossing transitions occur at interlayer separations $d/l$ marked by vertical dashed lines. The ground state in each region is indicated on the figures. (We note that due to the high number of possible crystal states at $\nu=$ 1/5, 39 in total, we have only plotted those with the most competitive energies.)

The richness of the bilayer phase diagram is evident. At $\nu=2/5$, the states that we find to be realized are $(3,3|\ 2)$, CS(2,1), $(4,4|\ 1)$, BG(2,0), and $(5,5|\ 0)$. At $\nu=1/3$, $(3,3|\ 3)$, CS(2,2), CS(2,1), $(5,5|\ 1)$, and BG(4,0) are realized. At $\nu=1/2$ the phase diagram is the same as that found by Scarola and Jain\cite{Scarola01b} with no crystal states. At $\nu=1/5$, we see the polarized FQH liquid $(5,5|\ 5)$, followed by a series of crystals with different symmetries, flavors of composite fermions and number of interlayer zeroes. 

Many features of the phase diagram are consistent with our expectation.

\begin{itemize}

\item In the limit of $d/l=0$, we obtain $(3,3|\ 2)$, $(3,3|\ 3)$, $(2,2|\ 2)$ and $(5,5|\ 5)$ states at $\nu=2/5$, 1/3, 1/2, and 1/5. With mapping to the single layer spinful system, these correspond to spin singlet 2/5, fully spin polarized 1/3, spin singlet 1/2, and fully spin polarized 1/5, which are known to be the lowest energy states. 

\item As expected, the integer $m$, which represents the strength of the interlayer correlations, decreases with increasing $d/l$.

\item The state in the limit of large $d/l$ is also consistent with our expectation. For $\nu=2/5$ we get two uncorrelated 1/5 states, and at $\nu=1/2$ two uncorrelated 1/4 CF Fermi seas. At $\nu=1/3$ and $\nu=1/5$, each layer has a triangular CF crystal, as expected for the individual layer fillings of $\nu=1/6$ and $\nu=1/10$, but these crystals are correlated into a BG crystal. The former is a $^4$CF crystal and the latter a $^6$CF crystal, as expected from previous calculations\cite{Chang05,Archer13}. 

\item For the total filling $\nu=1/2$, no crystal is stabilized according to our calculations. However, we note that the energy of the crystal BG$(2,0)$ is very close (within $0.002 e^2/\epsilon l$) to that of the independent layer state $(4\ 4|\ 0)$ in the limit of large separation.

\item 
At total filling $\nu=1/5$ we see that a crystal state appears quickly as we increase $d/l$. 
We see a large number of crystal-to-crystal transitions, and achieve each of the three crystal lattices that we have considered.
We note here that for states at this filling factor, the estimated error in the thermodynamic limit increases significantly, making it difficult to precisely ascertain the value of $d/l$ where the transition into the BG(6,0) crystal takes place.

\end{itemize}

We thus find a rich phase diagram of liquids and crystals resulting from tuning the the relative strengths of the intra-layer and interlayer interactions. Each filling factor considered here has its own complex evolution as the interlayer interaction is weakened.

\subsection{Finite Width}

We next consider the effects of finite width by looking at the same set of parameters for an effective
Coulomb potential in several double well geometries.

\begin{figure}
\includegraphics[width=3.8in]{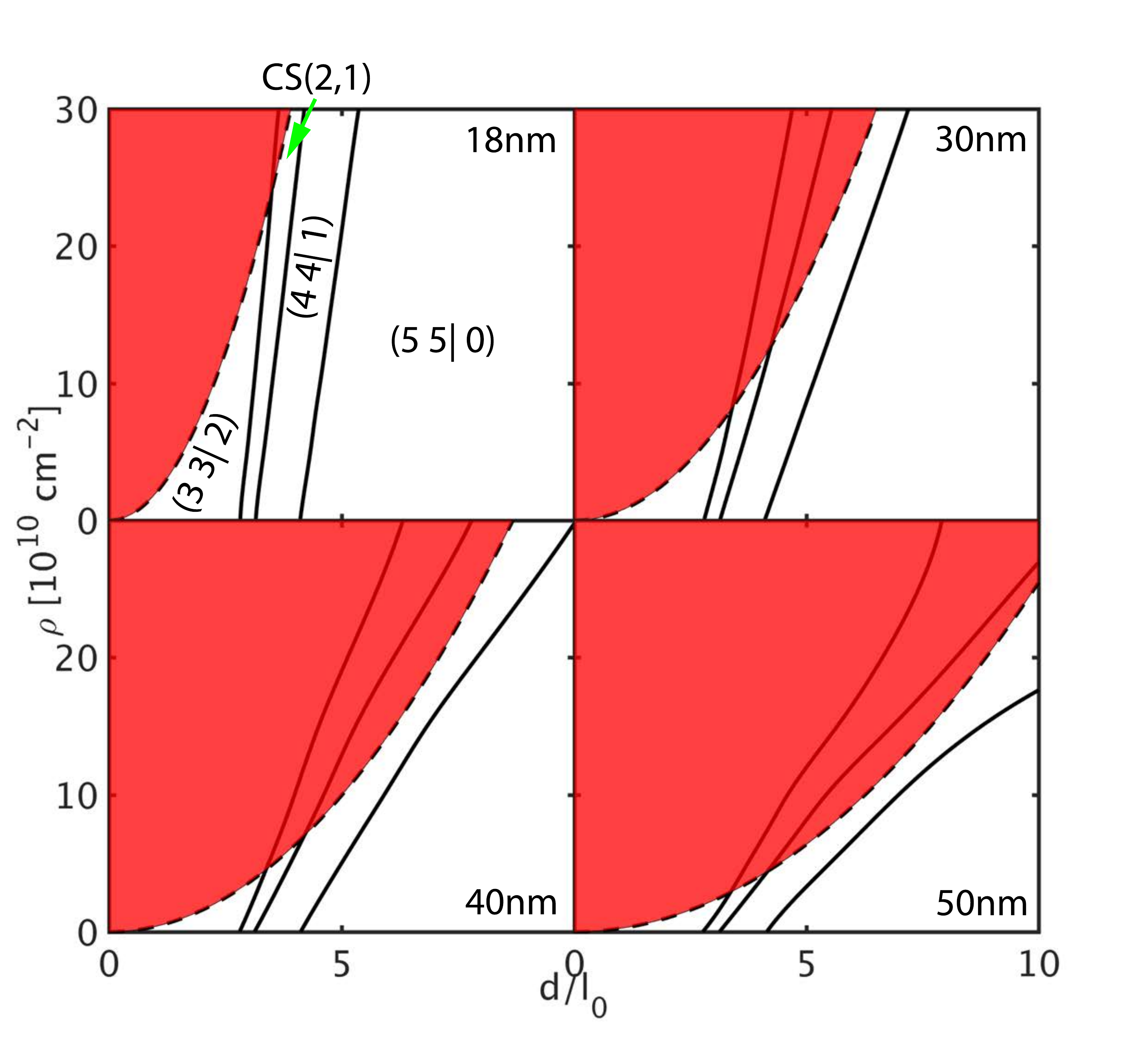}
\caption{Finite width phase diagram for $\nu = \frac{2}{5}$. We plot the phases expected in DQWs with individual well widths 18nm, 30nm, 40nm, and 50nm as a function of carrier density in units of 10$^{10}$ cm$^{-2}$ and layer separation in units of magnetic length. The shaded area is unphysical, as here the quantum well width exceeds the layer separation.  The phase diagram is qualitatively similar to that for zero width, except for the absence of the binary graphene crystal phase. 
}
\label{25fw}
\end{figure}

\begin{figure}
\includegraphics[width=3.8in]{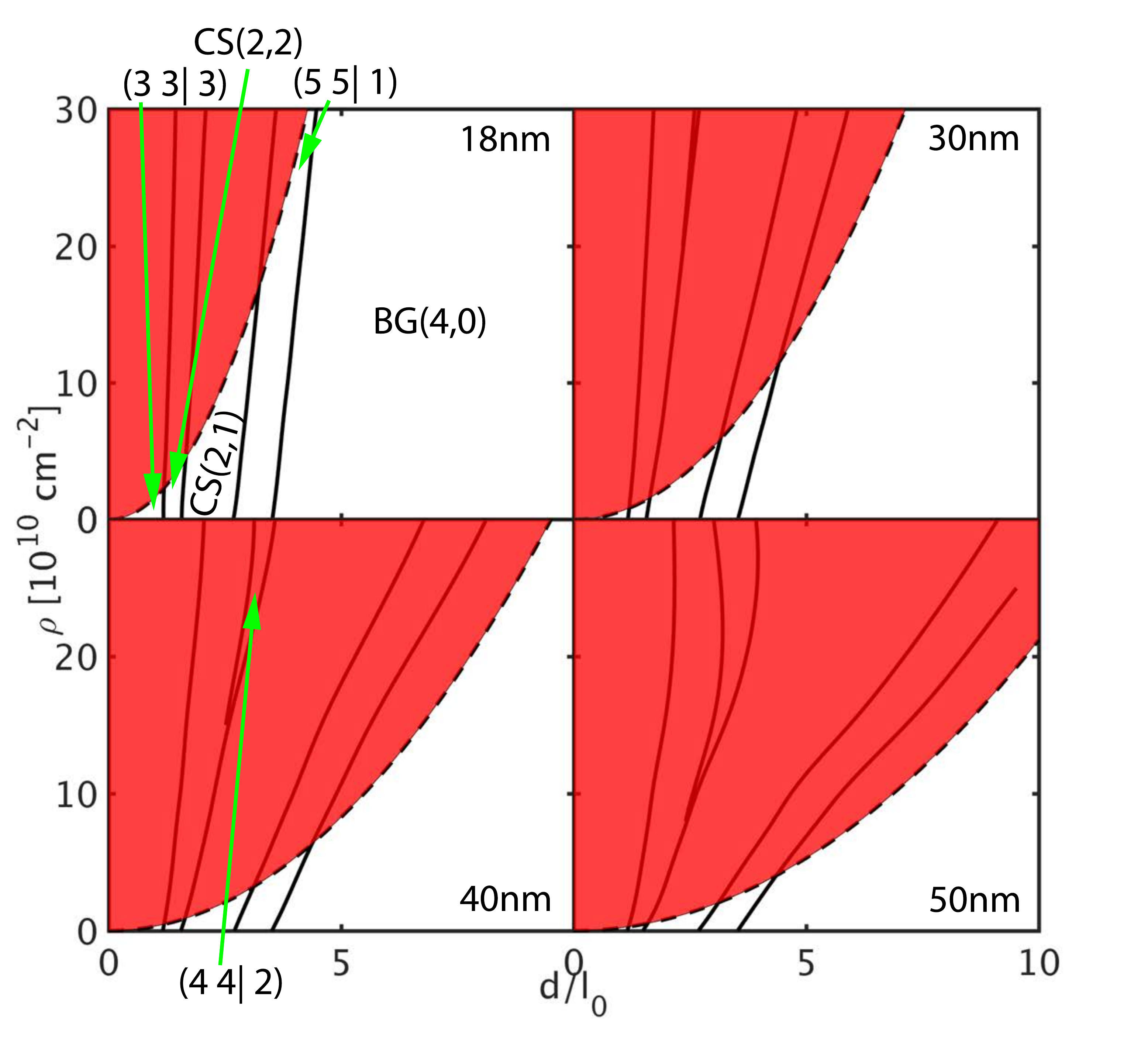}
\caption{Same as in Fig.~\ref{25fw} but for $\nu = \frac{1}{3}$. The phase diagram is qualitatively similar to that for zero width. 
}
\label{13fw}
\end{figure}

\begin{figure}
\includegraphics[width=3.8in]{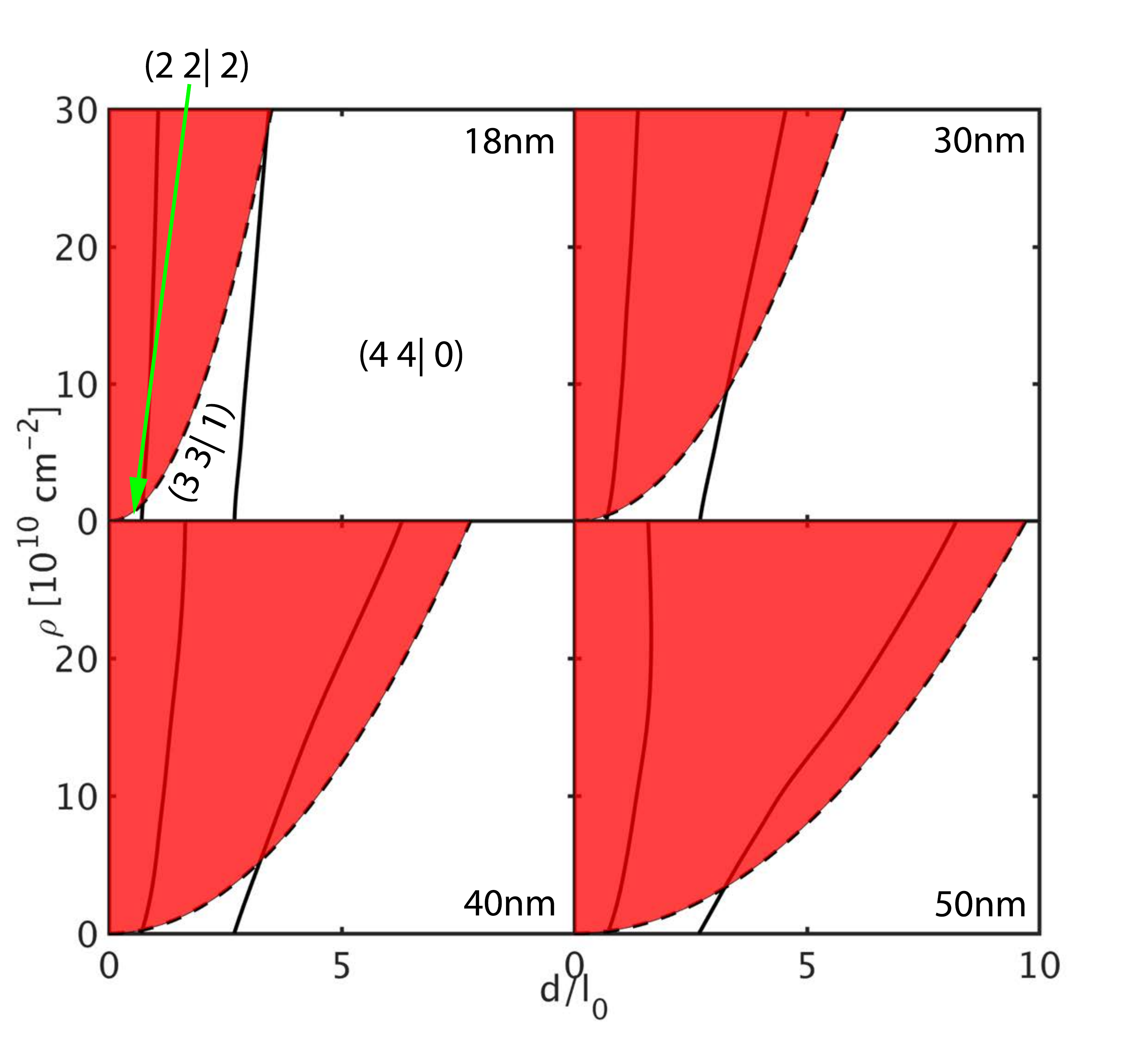}
\caption{Same as in Fig.~\ref{25fw} but for $\nu = \frac{1}{2}$. The phase diagram is qualitatively similar to that for zero width.
}
\label{12fw}
\end{figure}

\begin{figure}
\includegraphics[width = 3.8in]{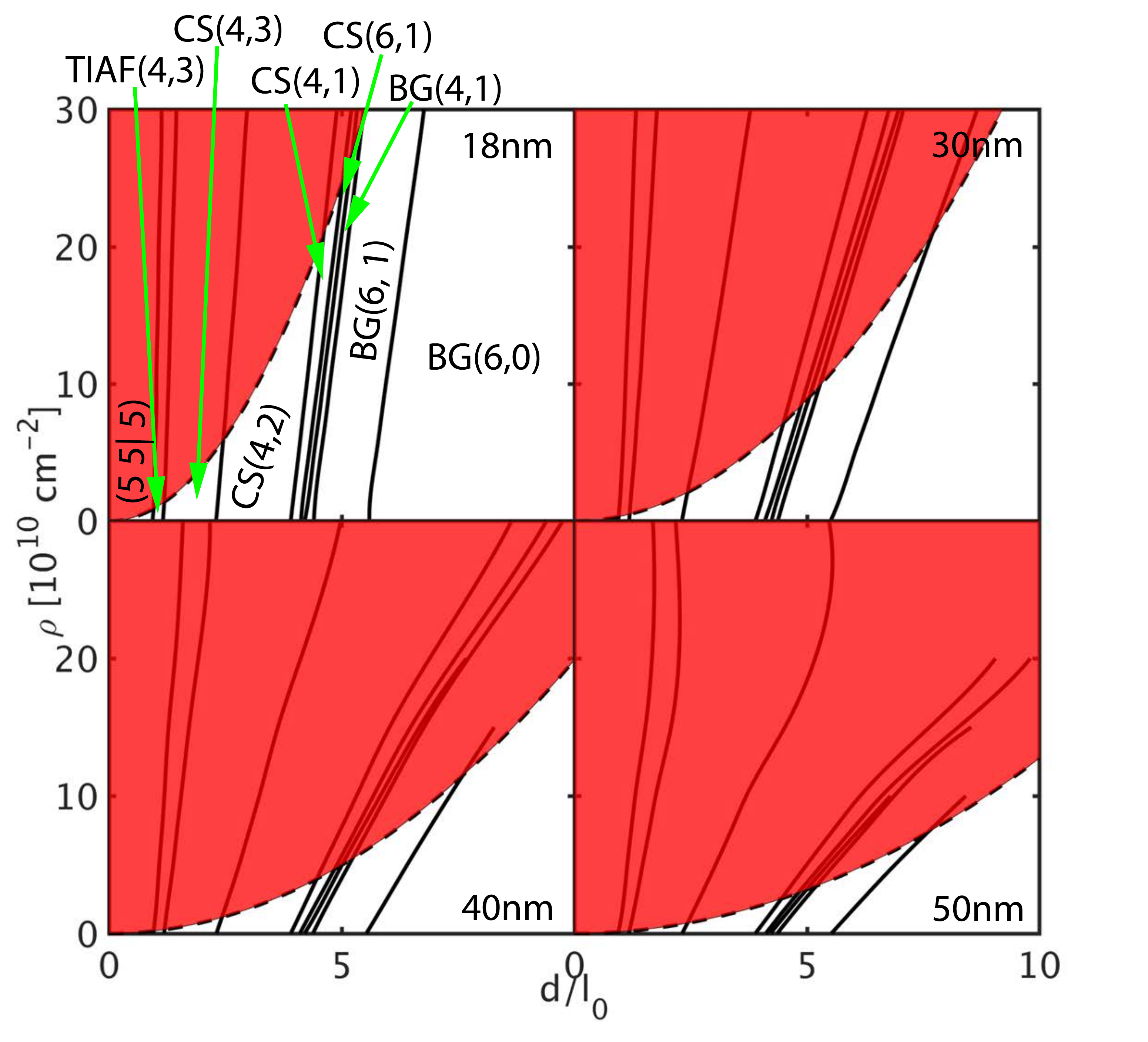}
\caption{Same as in Fig.~\ref{25fw} but for $\nu = \frac{1}{5}$. Here various crystal phases dominate the phase diagram. The phase diagram is qualitatively similar to that for zero width.
}
\label{15fw}
\end{figure}

In our finite width calculations, we consider double quantum well geometries with well widths of 18nm, 30nm, 40nm, and 50nm. The bilayer separation $d$ is taken as the center-to-center distance.
The finite width effects serve to alter the values of the separation at which the phase transitions occur, typically not changing the ordering of the states. Figures \ref{25fw}-\ref{15fw} show the phase diagrams for various widths for each filling in the $\rho$ - $d/l$ plane, where $\rho$ is the electron density. It is important to note that the region with $w>d$ is unphysical (two wells overlap) and has been shaded red. For each filling factor, the states are labeled only in the case of 18nm well width because the ordering of states for larger well widths is the same. 
We have not considered tunneling between layers, which may be important for small $d/l$ or for the bilayer interpretation with wide quantum wells.

We find that the ordering of states at each filling factor does not drastically change from that found for $d=0$. At filling factors 1/2, 1/3, and 1/5, we obtain the same states with the same ordering as for a zero width bilayer in the physical (unshaded) region.  
(Any differences from the zero width phase diagrams occur in the red shaded unphysical region.)
For filling factor 2/5, we find that the binary graphene phase is present in a narrow range for zero width, but is suppressed when we consider the finite width interaction.

\section{Comparison with experiment}

The results presented in this work apply to double quantum wells studied by Eisenstein {\em et al.}\cite{Eisenstein92}. These authors find an incompressible state at total filling $\nu=1/2$ in bilayers of quantum wells of width 18nm each for separations $d/l\approx 2.4-2.9$. That is consistent with our phase diagram for $\nu=1/2$. For separation $d/l=3.6$ they find an insulator, whereas in our phase diagram, the system with density 1.3$\times 10^{11}$ cm$^{-2}$ at $d/l=3.6$ is predicted to lie in the compressible phase $(4,4|0)$ which consists of two uncorrelated $1/4$ CF Fermi seas in each layer. There is no doubt that the $(4,4,|0)$ phase must ideally occur in the $d/l>>1$ limit, and therefore it is tempting to attribute the experimental insulating phase here to disorder. We suspect that disorder (enhanced due to the thin AlAs barrier layer) either freezes out the CF Fermi seas or stabilizes a bilayer crystal phase, which in this case would be a bilayer graphene crystal of composite fermions whose energy is very close to that of the $(4,4,|0)$ compressible state. A reliable account of disorder is outside the scope of our current study, but we note that disorder is expected to favor the crystal phase, which can accommodate disorder more readily than an incompressible liquid phase. In this context, it is also worth recalling that a crystal is often more competitive slightly away from the special fillings (an example being $\nu=1/5$ in a single layer system), and thus can swamp an incompressible FQH state in the presence of significant density inhomogeneities.

A direct comparison of our studies with the experimental results of Manoharan {\em et al.} and Hatke {\em et al.} \cite{Manoharan96, Hatke15} in WQW is not possible. In a WQW, the two ``layers" correspond to  even and odd combinations of the lowest symmetric and antisymmetric subbands, which are separated by a gap $\Delta_{\rm SAS}$, with the system making a transition from single layer-like at large $\Delta_{\rm SAS}$ to bilayer-like at small $\Delta_{\rm SAS}$. This system is akin to a bilayer with interlayer tunneling, which we have not considered in our paper. However, we can hope for a qualitative comparison, because in the bilayer-like region reducing $\Delta_{\rm SAS}$ is qualitatively similar to increasing the layer separation $d$. We list certain similarities and differences between previous WQW results and our predictions.

At $\nu = 2/5$, our calculations do not find a wide region of insulating phase that is present in WQW. In addition, we only find one crystal state when we consider finite width interactions as opposed to two crystals suggested by microwave spectroscopy measurements. At $\nu = 1/3$, we predict a reentrant incompressible FQH phase at intermediate separations, not seen in WQWs. We do find two separate crystals, correlated square and binary graphene, consistent with the transitions seen in microwave spectroscopy. At $\nu = 1/5$, we find the crystal phase to dominate the phase diagram, in qualitative agreement with the WQW experiment which finds a crystal phase immediately upon transition into a bilayer phase.  For filling $\nu = 1/2$, we do not find any crystal states to be stabilized, which is at odds with insulating behavior seen in WQW experiments. Again, disorder may be playing an important role in stabilizing some of the insulating phases. 

We note here that Thiebaut, Regnault and Goerbig \cite{Thiebaut15} have studied the WQW system at $\nu=1/2$ in the Hartree-Fock approximation. They find that a single layer crystal state occupying the second subband is stabilized for a parameter range that is in good agreement with the experimental phase diagram of Shabani {\em et al.}\cite{Shabani13}. 

In summary, we have performed a comprehensive study of both crystal and liquid phases in a bilayer system and obtained phase diagrams at several filling factors. In addition to the incompressible and compressible CF liquids, the phase diagrams also contain three types of CF crystals, namely Triangular Ising Antiferromagnet, Correlated Square, and Binary Graphene. We find that in addition to liquid-to-liquid transitions and liquid-to-crystal transitions, there are several crystal-to-crystal transitions in which the CF lattice reorders itself. We have made preliminary comparisons with existing experiments, and hope that this work will motivate a more systematic study of the insulating states in bilayer systems. 

{\bf Acknowledgments:} 
We thank Lloyd Engel, Mark Goerbig, Mansour Shayegan, Yuhe Zhang and Jianyun Zhao for insightful conversations. This work was supported in part by the US National Science Foundation, Grant No. DMR-1401636.


\begin{thebibliography}{83}
\expandafter\ifx\csname natexlab\endcsname\relax\def\natexlab#1{#1}\fi
\expandafter\ifx\csname bibnamefont\endcsname\relax
  \def\bibnamefont#1{#1}\fi
\expandafter\ifx\csname bibfnamefont\endcsname\relax
  \def\bibfnamefont#1{#1}\fi
\expandafter\ifx\csname citenamefont\endcsname\relax
  \def\citenamefont#1{#1}\fi
\expandafter\ifx\csname url\endcsname\relax
  \def\url#1{\texttt{#1}}\fi
\expandafter\ifx\csname urlprefix\endcsname\relax\def\urlprefix{URL }\fi
\providecommand{\bibinfo}[2]{#2}
\providecommand{\eprint}[2][]{\url{#2}}

\bibitem[{\citenamefont{Wigner}(1934)}]{Wigner34}
\bibinfo{author}{\bibfnamefont{E.}~\bibnamefont{Wigner}},
  \bibinfo{journal}{Phys. Rev.} \textbf{\bibinfo{volume}{46}},
  \bibinfo{pages}{1002} (\bibinfo{year}{1934}).

\bibitem[{\citenamefont{Lozovik and Yudson}(1975)}]{Lozovik75}
\bibinfo{author}{\bibfnamefont{Y.~E.} \bibnamefont{Lozovik}} \bibnamefont{and}
  \bibinfo{author}{\bibfnamefont{V.~I.} \bibnamefont{Yudson}},
  \bibinfo{journal}{JETP Lett.} \textbf{\bibinfo{volume}{22}},
  \bibinfo{pages}{(11), 274-276} (\bibinfo{year}{1975}).

\bibitem[{\citenamefont{Tsui et~al.}(1982)\citenamefont{Tsui, Stormer, and
  Gossard}}]{Tsui82}
\bibinfo{author}{\bibfnamefont{D.~C.} \bibnamefont{Tsui}},
  \bibinfo{author}{\bibfnamefont{H.~L.} \bibnamefont{Stormer}},
  \bibnamefont{and} \bibinfo{author}{\bibfnamefont{A.~C.}
  \bibnamefont{Gossard}}, \bibinfo{journal}{Phys. Rev. Lett.}
  \textbf{\bibinfo{volume}{48}}, \bibinfo{pages}{1559} (\bibinfo{year}{1982}),
  \urlprefix\url{http://link.aps.org/doi/10.1103/PhysRevLett.48.1559}.

\bibitem[{\citenamefont{Lam and Girvin}(1984)}]{Lam84}
\bibinfo{author}{\bibfnamefont{P.~K.} \bibnamefont{Lam}} \bibnamefont{and}
  \bibinfo{author}{\bibfnamefont{S.~M.} \bibnamefont{Girvin}},
  \bibinfo{journal}{Phys. Rev. B} \textbf{\bibinfo{volume}{30}},
  \bibinfo{pages}{473} (\bibinfo{year}{1984}).

\bibitem[{\citenamefont{Levesque et~al.}(1984)\citenamefont{Levesque, Weis, and
  MacDonald}}]{Levesque84}
\bibinfo{author}{\bibfnamefont{D.}~\bibnamefont{Levesque}},
  \bibinfo{author}{\bibfnamefont{J.~J.} \bibnamefont{Weis}}, \bibnamefont{and}
  \bibinfo{author}{\bibfnamefont{A.~H.} \bibnamefont{MacDonald}},
  \bibinfo{journal}{Phys. Rev. B} \textbf{\bibinfo{volume}{30}},
  \bibinfo{pages}{1056} (\bibinfo{year}{1984}).

\bibitem[{\citenamefont{Esfarjani and Chui}(1990)}]{Esfarjani90}
\bibinfo{author}{\bibfnamefont{K.}~\bibnamefont{Esfarjani}} \bibnamefont{and}
  \bibinfo{author}{\bibfnamefont{S.~T.} \bibnamefont{Chui}},
  \bibinfo{journal}{Phys. Rev. B} \textbf{\bibinfo{volume}{42}},
  \bibinfo{pages}{10758} (\bibinfo{year}{1990}).

\bibitem[{\citenamefont{C\^ot\'e and MacDonald}(1991)}]{Cote91}
\bibinfo{author}{\bibfnamefont{R.}~\bibnamefont{C\^ot\'e}} \bibnamefont{and}
  \bibinfo{author}{\bibfnamefont{A.~H.} \bibnamefont{MacDonald}},
  \bibinfo{journal}{Phys. Rev. B} \textbf{\bibinfo{volume}{44}},
  \bibinfo{pages}{8759} (\bibinfo{year}{1991}).

\bibitem[{\citenamefont{Zheng and Fertig}(1995)}]{Zheng95}
\bibinfo{author}{\bibfnamefont{L.}~\bibnamefont{Zheng}} \bibnamefont{and}
  \bibinfo{author}{\bibfnamefont{H.~A.} \bibnamefont{Fertig}},
  \bibinfo{journal}{Phys. Rev. B} \textbf{\bibinfo{volume}{52}},
  \bibinfo{pages}{12282} (\bibinfo{year}{1995}),
  \urlprefix\url{https://link.aps.org/doi/10.1103/PhysRevB.52.12282}.

\bibitem[{\citenamefont{Goldoni and Peeters}(1996)}]{Goldoni96}
\bibinfo{author}{\bibfnamefont{G.}~\bibnamefont{Goldoni}} \bibnamefont{and}
  \bibinfo{author}{\bibfnamefont{F.~M.} \bibnamefont{Peeters}},
  \bibinfo{journal}{Phys. Rev. B} \textbf{\bibinfo{volume}{53}},
  \bibinfo{pages}{4591} (\bibinfo{year}{1996}),
  \urlprefix\url{https://link.aps.org/doi/10.1103/PhysRevB.53.4591}.

\bibitem[{\citenamefont{Yang et~al.}(2001)\citenamefont{Yang, Haldane, and
  Rezayi}}]{Yang01}
\bibinfo{author}{\bibfnamefont{K.}~\bibnamefont{Yang}},
  \bibinfo{author}{\bibfnamefont{F.~D.~M.} \bibnamefont{Haldane}},
  \bibnamefont{and} \bibinfo{author}{\bibfnamefont{E.~H.}
  \bibnamefont{Rezayi}}, \bibinfo{journal}{Phys. Rev. B}
  \textbf{\bibinfo{volume}{64}}, \bibinfo{pages}{081301}
  (\bibinfo{year}{2001}).

\bibitem[{\citenamefont{Shibata and Yoshioka}(2003)}]{Shibata03}
\bibinfo{author}{\bibfnamefont{N.}~\bibnamefont{Shibata}} \bibnamefont{and}
  \bibinfo{author}{\bibfnamefont{D.}~\bibnamefont{Yoshioka}}, \bibinfo{journal}{J. Phys. Soc. Jpn.}
  \textbf{\bibinfo{volume}{72}}, \bibinfo{pages}{664} (\bibinfo{year}{2003}).

\bibitem[{\citenamefont{He et~al.}(2005)\citenamefont{He, Cui, Ma, Chen, Liu,
  and Zou}}]{He05}
\bibinfo{author}{\bibfnamefont{W.~J.} \bibnamefont{He}},
  \bibinfo{author}{\bibfnamefont{T.}~\bibnamefont{Cui}},
  \bibinfo{author}{\bibfnamefont{Y.~M.} \bibnamefont{Ma}},
  \bibinfo{author}{\bibfnamefont{C.~B.} \bibnamefont{Chen}},
  \bibinfo{author}{\bibfnamefont{Z.~M.} \bibnamefont{Liu}}, \bibnamefont{and}
  \bibinfo{author}{\bibfnamefont{G.~T.} \bibnamefont{Zou}},
  \bibinfo{journal}{Phys. Rev. B} \textbf{\bibinfo{volume}{72}},
  \bibinfo{pages}{195306} (\bibinfo{year}{2005}).

\bibitem[{\citenamefont{Shi et~al.}(2007)\citenamefont{Shi, Jeon, and
  Jain}}]{Shi07}
\bibinfo{author}{\bibfnamefont{C.}~\bibnamefont{Shi}},
  \bibinfo{author}{\bibfnamefont{G.~S.} \bibnamefont{Jeon}}, \bibnamefont{and}
  \bibinfo{author}{\bibfnamefont{J.~K.} \bibnamefont{Jain}},
  \bibinfo{journal}{Phys. Rev. B} \textbf{\bibinfo{volume}{75}},
  \bibinfo{pages}{165302} (\bibinfo{year}{2007}).

\bibitem[{\citenamefont{Goldman et~al.}(1990)\citenamefont{Goldman, Santos,
  Shayegan, and Cunningham}}]{Goldman90}
\bibinfo{author}{\bibfnamefont{V.~J.} \bibnamefont{Goldman}},
  \bibinfo{author}{\bibfnamefont{M.}~\bibnamefont{Santos}},
  \bibinfo{author}{\bibfnamefont{M.}~\bibnamefont{Shayegan}}, \bibnamefont{and}
  \bibinfo{author}{\bibfnamefont{J.~E.} \bibnamefont{Cunningham}},
  \bibinfo{journal}{Phys. Rev. Lett.} \textbf{\bibinfo{volume}{65}},
  \bibinfo{pages}{2189} (\bibinfo{year}{1990}).

\bibitem[{\citenamefont{Jiang et~al.}(1990)\citenamefont{Jiang, Willett,
  Stormer, Tsui, Pfeiffer, and West}}]{Jiang90}
\bibinfo{author}{\bibfnamefont{H.~W.} \bibnamefont{Jiang}},
  \bibinfo{author}{\bibfnamefont{R.~L.} \bibnamefont{Willett}},
  \bibinfo{author}{\bibfnamefont{H.~L.} \bibnamefont{Stormer}},
  \bibinfo{author}{\bibfnamefont{D.~C.} \bibnamefont{Tsui}},
  \bibinfo{author}{\bibfnamefont{L.~N.} \bibnamefont{Pfeiffer}},
  \bibnamefont{and} \bibinfo{author}{\bibfnamefont{K.~W.} \bibnamefont{West}},
  \bibinfo{journal}{Phys. Rev. Lett.} \textbf{\bibinfo{volume}{65}},
  \bibinfo{pages}{633} (\bibinfo{year}{1990}).

\bibitem[{\citenamefont{Jiang et~al.}(1991)\citenamefont{Jiang, Stormer, Tsui,
  Pfeiffer, and West}}]{Jiang91}
\bibinfo{author}{\bibfnamefont{H.~W.} \bibnamefont{Jiang}},
  \bibinfo{author}{\bibfnamefont{H.~L.} \bibnamefont{Stormer}},
  \bibinfo{author}{\bibfnamefont{D.~C.} \bibnamefont{Tsui}},
  \bibinfo{author}{\bibfnamefont{L.~N.} \bibnamefont{Pfeiffer}},
  \bibnamefont{and} \bibinfo{author}{\bibfnamefont{K.~W.} \bibnamefont{West}},
  \bibinfo{journal}{Phys. Rev. B} \textbf{\bibinfo{volume}{44}},
  \bibinfo{pages}{8107} (\bibinfo{year}{1991}).

\bibitem[{\citenamefont{Williams et~al.}(1991)\citenamefont{Williams, Wright,
  Clark, Andrei, Deville, Glattli, Probst, Etienne, Dorin, Foxon
  et~al.}}]{Williams91}
\bibinfo{author}{\bibfnamefont{F.~I.~B.} \bibnamefont{Williams}},
  \bibinfo{author}{\bibfnamefont{P.~A.} \bibnamefont{Wright}},
  \bibinfo{author}{\bibfnamefont{R.~G.} \bibnamefont{Clark}},
  \bibinfo{author}{\bibfnamefont{E.~Y.} \bibnamefont{Andrei}},
  \bibinfo{author}{\bibfnamefont{G.}~\bibnamefont{Deville}},
  \bibinfo{author}{\bibfnamefont{D.~C.} \bibnamefont{Glattli}},
  \bibinfo{author}{\bibfnamefont{O.}~\bibnamefont{Probst}},
  \bibinfo{author}{\bibfnamefont{B.}~\bibnamefont{Etienne}},
  \bibinfo{author}{\bibfnamefont{C.}~\bibnamefont{Dorin}},
  \bibinfo{author}{\bibfnamefont{C.~T.} \bibnamefont{Foxon}},
  \bibnamefont{et~al.}, \bibinfo{journal}{Phys. Rev. Lett.}
  \textbf{\bibinfo{volume}{66}}, \bibinfo{pages}{3285} (\bibinfo{year}{1991}),
  \urlprefix\url{https://link.aps.org/doi/10.1103/PhysRevLett.66.3285}.

\bibitem[{\citenamefont{Manoharan and Shayegan}(1994)}]{Manoharan94a}
\bibinfo{author}{\bibfnamefont{H.~C.} \bibnamefont{Manoharan}}
  \bibnamefont{and} \bibinfo{author}{\bibfnamefont{M.}~\bibnamefont{Shayegan}},
  \bibinfo{journal}{Phys. Rev. B} \textbf{\bibinfo{volume}{50}},
  \bibinfo{pages}{17662} (\bibinfo{year}{1994}).

\bibitem[{\citenamefont{Pan et~al.}(2002)\citenamefont{Pan, Stormer, Tsui,
  Pfeiffer, Baldwin, and West}}]{Pan02}
\bibinfo{author}{\bibfnamefont{W.}~\bibnamefont{Pan}},
  \bibinfo{author}{\bibfnamefont{H.~L.} \bibnamefont{Stormer}},
  \bibinfo{author}{\bibfnamefont{D.~C.} \bibnamefont{Tsui}},
  \bibinfo{author}{\bibfnamefont{L.~N.} \bibnamefont{Pfeiffer}},
  \bibinfo{author}{\bibfnamefont{K.~W.} \bibnamefont{Baldwin}},
  \bibnamefont{and} \bibinfo{author}{\bibfnamefont{K.~W.} \bibnamefont{West}},
  \bibinfo{journal}{Phys. Rev. Lett.} \textbf{\bibinfo{volume}{88}},
  \bibinfo{pages}{176802} (\bibinfo{year}{2002}).

\bibitem[{\citenamefont{Engel et~al.}(1997)\citenamefont{Engel, Li, Shahar,
  Tsui, and Shayegan}}]{Engel97}
\bibinfo{author}{\bibfnamefont{L.}~\bibnamefont{Engel}},
  \bibinfo{author}{\bibfnamefont{C.-C.} \bibnamefont{Li}},
  \bibinfo{author}{\bibfnamefont{D.}~\bibnamefont{Shahar}},
  \bibinfo{author}{\bibfnamefont{D.}~\bibnamefont{Tsui}}, \bibnamefont{and}
  \bibinfo{author}{\bibfnamefont{M.}~\bibnamefont{Shayegan}},
  \bibinfo{journal}{Physica E} \textbf{\bibinfo{volume}{1}},
  \bibinfo{pages}{111 } (\bibinfo{year}{1997}).

\bibitem[{\citenamefont{Li et~al.}(2000)\citenamefont{Li, Yoon, Engel, Shahar,
  Tsui, and Shayegan}}]{Li00}
\bibinfo{author}{\bibfnamefont{C.-C.} \bibnamefont{Li}},
  \bibinfo{author}{\bibfnamefont{J.}~\bibnamefont{Yoon}},
  \bibinfo{author}{\bibfnamefont{L.~W.} \bibnamefont{Engel}},
  \bibinfo{author}{\bibfnamefont{D.}~\bibnamefont{Shahar}},
  \bibinfo{author}{\bibfnamefont{D.~C.} \bibnamefont{Tsui}}, \bibnamefont{and}
  \bibinfo{author}{\bibfnamefont{M.}~\bibnamefont{Shayegan}},
  \bibinfo{journal}{Phys. Rev. B} \textbf{\bibinfo{volume}{61}},
  \bibinfo{pages}{10905} (\bibinfo{year}{2000}).

\bibitem[{\citenamefont{Ye et~al.}(2002)\citenamefont{Ye, Engel, Tsui, Lewis,
  Pfeiffer, and West}}]{Ye02}
\bibinfo{author}{\bibfnamefont{P.~D.} \bibnamefont{Ye}},
  \bibinfo{author}{\bibfnamefont{L.~W.} \bibnamefont{Engel}},
  \bibinfo{author}{\bibfnamefont{D.~C.} \bibnamefont{Tsui}},
  \bibinfo{author}{\bibfnamefont{R.~M.} \bibnamefont{Lewis}},
  \bibinfo{author}{\bibfnamefont{L.~N.} \bibnamefont{Pfeiffer}},
  \bibnamefont{and} \bibinfo{author}{\bibfnamefont{K.}~\bibnamefont{West}},
  \bibinfo{journal}{Phys. Rev. Lett.} \textbf{\bibinfo{volume}{89}},
  \bibinfo{pages}{176802} (\bibinfo{year}{2002}).

\bibitem[{\citenamefont{Chen et~al.}(2004{\natexlab{a}})\citenamefont{Chen,
  Lewis, Engel, Tsui, Ye, Wang, Pfeiffer, and West}}]{Chen04}
\bibinfo{author}{\bibfnamefont{Y.~P.} \bibnamefont{Chen}},
  \bibinfo{author}{\bibfnamefont{R.~M.} \bibnamefont{Lewis}},
  \bibinfo{author}{\bibfnamefont{L.~W.} \bibnamefont{Engel}},
  \bibinfo{author}{\bibfnamefont{D.~C.} \bibnamefont{Tsui}},
  \bibinfo{author}{\bibfnamefont{P.~D.} \bibnamefont{Ye}},
  \bibinfo{author}{\bibfnamefont{Z.~H.} \bibnamefont{Wang}},
  \bibinfo{author}{\bibfnamefont{L.~N.} \bibnamefont{Pfeiffer}},
  \bibnamefont{and} \bibinfo{author}{\bibfnamefont{K.~W.} \bibnamefont{West}},
  \bibinfo{journal}{Phys. Rev. Lett.} \textbf{\bibinfo{volume}{93}},
  \bibinfo{pages}{206805} (\bibinfo{year}{2004}{\natexlab{b}}),
  \urlprefix\url{https://link.aps.org/doi/10.1103/PhysRevLett.93.206805}.

\bibitem[{\citenamefont{Sambandamurthy
  et~al.}(2006)\citenamefont{Sambandamurthy, Wang, Lewis, Chen, Engel, Tsui,
  Pfeiffer, and West}}]{Sambandamurthy06}
\bibinfo{author}{\bibfnamefont{G.}~\bibnamefont{Sambandamurthy}},
  \bibinfo{author}{\bibfnamefont{Z.}~\bibnamefont{Wang}},
  \bibinfo{author}{\bibfnamefont{R.}~\bibnamefont{Lewis}},
  \bibinfo{author}{\bibfnamefont{Y.~P.} \bibnamefont{Chen}},
  \bibinfo{author}{\bibfnamefont{L.}~\bibnamefont{Engel}},
  \bibinfo{author}{\bibfnamefont{D.}~\bibnamefont{Tsui}},
  \bibinfo{author}{\bibfnamefont{L.}~\bibnamefont{Pfeiffer}}, \bibnamefont{and}
  \bibinfo{author}{\bibfnamefont{K.}~\bibnamefont{West}},
  \bibinfo{journal}{Solid State Commun.} \textbf{\bibinfo{volume}{140}},
  \bibinfo{pages}{100 } (\bibinfo{year}{2006}).

\bibitem[{\citenamefont{Chen et~al.}(2006)\citenamefont{Chen, Sambandamurthy,
  Wang, Lewis, Engel, Tsui, Ye, Pfeiffer, and West}}]{Chen06}
\bibinfo{author}{\bibfnamefont{Y.~P.} \bibnamefont{Chen}},
  \bibinfo{author}{\bibfnamefont{G.}~\bibnamefont{Sambandamurthy}},
  \bibinfo{author}{\bibfnamefont{Z.~H.} \bibnamefont{Wang}},
  \bibinfo{author}{\bibfnamefont{R.~M.} \bibnamefont{Lewis}},
  \bibinfo{author}{\bibfnamefont{L.~W.} \bibnamefont{Engel}},
  \bibinfo{author}{\bibfnamefont{D.~C.} \bibnamefont{Tsui}},
  \bibinfo{author}{\bibfnamefont{P.~D.} \bibnamefont{Ye}},
  \bibinfo{author}{\bibfnamefont{L.~N.} \bibnamefont{Pfeiffer}},
  \bibnamefont{and} \bibinfo{author}{\bibfnamefont{K.~W.} \bibnamefont{West}},
  \bibinfo{journal}{Nature Phys.} \textbf{\bibinfo{volume}{2}},
  \bibinfo{pages}{452} (\bibinfo{year}{2006}).

\bibitem[{\citenamefont{Yi and Fertig}(1998)}]{Yi98}
\bibinfo{author}{\bibfnamefont{H.}~\bibnamefont{Yi}} \bibnamefont{and}
  \bibinfo{author}{\bibfnamefont{H.~A.} \bibnamefont{Fertig}},
  \bibinfo{journal}{Phys. Rev. B} \textbf{\bibinfo{volume}{58}},
  \bibinfo{pages}{4019} (\bibinfo{year}{1998}).

\bibitem[{\citenamefont{Narevich et~al.}(2001)\citenamefont{Narevich, Murthy,
  and Fertig}}]{Narevich01}
\bibinfo{author}{\bibfnamefont{R.}~\bibnamefont{Narevich}},
  \bibinfo{author}{\bibfnamefont{G.}~\bibnamefont{Murthy}}, \bibnamefont{and}
  \bibinfo{author}{\bibfnamefont{H.~A.} \bibnamefont{Fertig}},
  \bibinfo{journal}{Phys. Rev. B} \textbf{\bibinfo{volume}{64}},
  \bibinfo{pages}{245326} (\bibinfo{year}{2001}).

\bibitem[{\citenamefont{Mandal et~al.}(2003)\citenamefont{Mandal, Peterson, and
  Jain}}]{Mandal03}
\bibinfo{author}{\bibfnamefont{S.~S.} \bibnamefont{Mandal}},
  \bibinfo{author}{\bibfnamefont{M.~R.} \bibnamefont{Peterson}},
  \bibnamefont{and} \bibinfo{author}{\bibfnamefont{J.~K.} \bibnamefont{Jain}},
  \bibinfo{journal}{Phys. Rev. Lett.} \textbf{\bibinfo{volume}{90}},
  \bibinfo{pages}{106403} (\bibinfo{year}{2003}).

\bibitem[{\citenamefont{Chang et~al.}(2005)\citenamefont{Chang, Jeon, and
  Jain}}]{Chang05}
\bibinfo{author}{\bibfnamefont{C.-C.} \bibnamefont{Chang}},
  \bibinfo{author}{\bibfnamefont{G.~S.} \bibnamefont{Jeon}}, \bibnamefont{and}
  \bibinfo{author}{\bibfnamefont{J.~K.} \bibnamefont{Jain}},
  \bibinfo{journal}{Phys. Rev. Lett.} \textbf{\bibinfo{volume}{94}},
  \bibinfo{pages}{016809} (\bibinfo{year}{2005}).

\bibitem[{\citenamefont{Archer et~al.}(2013)\citenamefont{Archer, Park, and
  Jain}}]{Archer13}
\bibinfo{author}{\bibfnamefont{A.~C.} \bibnamefont{Archer}},
  \bibinfo{author}{\bibfnamefont{K.}~\bibnamefont{Park}}, \bibnamefont{and}
  \bibinfo{author}{\bibfnamefont{J.~K.} \bibnamefont{Jain}},
  \bibinfo{journal}{Phys. Rev. Lett.} \textbf{\bibinfo{volume}{111}},
  \bibinfo{pages}{146804} (\bibinfo{year}{2013}).

\bibitem[{\citenamefont{Liu et~al.}(2014)\citenamefont{Liu, Kamburov, Hasdemir,
  Shayegan, Pfeiffer, West, and Baldwin}}]{Liu14a}
\bibinfo{author}{\bibfnamefont{Y.}~\bibnamefont{Liu}},
  \bibinfo{author}{\bibfnamefont{D.}~\bibnamefont{Kamburov}},
  \bibinfo{author}{\bibfnamefont{S.}~\bibnamefont{Hasdemir}},
  \bibinfo{author}{\bibfnamefont{M.}~\bibnamefont{Shayegan}},
  \bibinfo{author}{\bibfnamefont{L.~N.} \bibnamefont{Pfeiffer}},
  \bibinfo{author}{\bibfnamefont{K.~W.} \bibnamefont{West}}, \bibnamefont{and}
  \bibinfo{author}{\bibfnamefont{K.~W.} \bibnamefont{Baldwin}},
  \bibinfo{journal}{Phys. Rev. Lett.} \textbf{\bibinfo{volume}{113}},
  \bibinfo{pages}{246803} (\bibinfo{year}{2014}),
  \urlprefix\url{http://link.aps.org/doi/10.1103/PhysRevLett.113.246803}.

\bibitem[{\citenamefont{Zhang et~al.}(2015)\citenamefont{Zhang, Du, Manfra,
  Pfeiffer, and West}}]{Zhang15c}
\bibinfo{author}{\bibfnamefont{C.}~\bibnamefont{Zhang}},
  \bibinfo{author}{\bibfnamefont{R.-R.} \bibnamefont{Du}},
  \bibinfo{author}{\bibfnamefont{M.~J.} \bibnamefont{Manfra}},
  \bibinfo{author}{\bibfnamefont{L.~N.} \bibnamefont{Pfeiffer}},
  \bibnamefont{and} \bibinfo{author}{\bibfnamefont{K.~W.} \bibnamefont{West}},
  \bibinfo{journal}{Phys. Rev. B} \textbf{\bibinfo{volume}{92}},
  \bibinfo{pages}{075434} (\bibinfo{year}{2015}),
  \urlprefix\url{https://link.aps.org/doi/10.1103/PhysRevB.92.075434}.

\bibitem[{\citenamefont{Jang et~al.}(2017)\citenamefont{Jang, Hunt, Pfeiffer,
  West, and Ashoori}}]{Jang17}
\bibinfo{author}{\bibfnamefont{J.}~\bibnamefont{Jang}},
  \bibinfo{author}{\bibfnamefont{B.~M.} \bibnamefont{Hunt}},
  \bibinfo{author}{\bibfnamefont{L.~N.} \bibnamefont{Pfeiffer}},
  \bibinfo{author}{\bibfnamefont{K.~W.} \bibnamefont{West}}, \bibnamefont{and}
  \bibinfo{author}{\bibfnamefont{R.~C.} \bibnamefont{Ashoori}},
  \bibinfo{journal}{Nature Physics} \textbf{\bibinfo{volume}{13}},
  \bibinfo{pages}{340} (\bibinfo{year}{2017}), ISSN \bibinfo{issn}{1745-2473}.

\bibitem[{\citenamefont{Chakraborty and Pietil\"ainen}(1987)}]{Chakraborty87}
\bibinfo{author}{\bibfnamefont{T.}~\bibnamefont{Chakraborty}} \bibnamefont{and}
  \bibinfo{author}{\bibfnamefont{P.}~\bibnamefont{Pietil\"ainen}},
  \bibinfo{journal}{Phys. Rev. Lett.} \textbf{\bibinfo{volume}{59}},
  \bibinfo{pages}{2784} (\bibinfo{year}{1987}),
  \urlprefix\url{https://link.aps.org/doi/10.1103/PhysRevLett.59.2784}.

\bibitem[{\citenamefont{Yoshioka et~al.}(1989)\citenamefont{Yoshioka,
  MacDonald, and Girvin}}]{Yoshioka89}
\bibinfo{author}{\bibfnamefont{D.}~\bibnamefont{Yoshioka}},
  \bibinfo{author}{\bibfnamefont{A.~H.} \bibnamefont{MacDonald}},
  \bibnamefont{and} \bibinfo{author}{\bibfnamefont{S.~M.}
  \bibnamefont{Girvin}}, \bibinfo{journal}{Phys. Rev. B}
  \textbf{\bibinfo{volume}{39}}, \bibinfo{pages}{1932} (\bibinfo{year}{1989}),
  \urlprefix\url{https://link.aps.org/doi/10.1103/PhysRevB.39.1932}.

\bibitem[{\citenamefont{He et~al.}(1991)\citenamefont{He, Xie, Das~Sarma, and
  Zhang}}]{He91}
\bibinfo{author}{\bibfnamefont{S.}~\bibnamefont{He}},
  \bibinfo{author}{\bibfnamefont{X.~C.} \bibnamefont{Xie}},
  \bibinfo{author}{\bibfnamefont{S.}~\bibnamefont{Das~Sarma}},
  \bibnamefont{and} \bibinfo{author}{\bibfnamefont{F.~C.} \bibnamefont{Zhang}},
  \bibinfo{journal}{Phys. Rev. B} \textbf{\bibinfo{volume}{43}},
  \bibinfo{pages}{9339} (\bibinfo{year}{1991}),
  \urlprefix\url{https://link.aps.org/doi/10.1103/PhysRevB.43.9339}.

\bibitem[{\citenamefont{He et~al.}(1993)\citenamefont{He, Das~Sarma, and
  Xie}}]{He93}
\bibinfo{author}{\bibfnamefont{S.}~\bibnamefont{He}},
  \bibinfo{author}{\bibfnamefont{S.}~\bibnamefont{Das~Sarma}},
  \bibnamefont{and} \bibinfo{author}{\bibfnamefont{X.~C.} \bibnamefont{Xie}},
  \bibinfo{journal}{Phys. Rev. B} \textbf{\bibinfo{volume}{47}},
  \bibinfo{pages}{4394} (\bibinfo{year}{1993}),
  \urlprefix\url{http://link.aps.org/doi/10.1103/PhysRevB.47.4394}.

\bibitem[{\citenamefont{Scarola and Jain}(2001)}]{Scarola01b}
\bibinfo{author}{\bibfnamefont{V.~W.} \bibnamefont{Scarola}} \bibnamefont{and}
  \bibinfo{author}{\bibfnamefont{J.~K.} \bibnamefont{Jain}},
  \bibinfo{journal}{Phys. Rev. B} \textbf{\bibinfo{volume}{64}},
  \bibinfo{pages}{085313} (\bibinfo{year}{2001}),
  \urlprefix\url{http://link.aps.org/doi/10.1103/PhysRevB.64.085313}.

\bibitem[{\citenamefont{Thiebaut et~al.}(2014)\citenamefont{Thiebaut, Goerbig,
  and Regnault}}]{Thiebaut14}
\bibinfo{author}{\bibfnamefont{N.}~\bibnamefont{Thiebaut}},
  \bibinfo{author}{\bibfnamefont{M.~O.} \bibnamefont{Goerbig}},
  \bibnamefont{and} \bibinfo{author}{\bibfnamefont{N.}~\bibnamefont{Regnault}},
  \bibinfo{journal}{Phys. Rev. B} \textbf{\bibinfo{volume}{89}},
  \bibinfo{pages}{195421} (\bibinfo{year}{2014}),
  \urlprefix\url{https://link.aps.org/doi/10.1103/PhysRevB.89.195421}.

\bibitem[{\citenamefont{Eisenstein et~al.}(1992)\citenamefont{Eisenstein,
  Boebinger, Pfeiffer, West, and He}}]{Eisenstein92}
\bibinfo{author}{\bibfnamefont{J.~P.} \bibnamefont{Eisenstein}},
  \bibinfo{author}{\bibfnamefont{G.~S.} \bibnamefont{Boebinger}},
  \bibinfo{author}{\bibfnamefont{L.~N.} \bibnamefont{Pfeiffer}},
  \bibinfo{author}{\bibfnamefont{K.~W.} \bibnamefont{West}}, \bibnamefont{and}
  \bibinfo{author}{\bibfnamefont{S.}~\bibnamefont{He}}, \bibinfo{journal}{Phys.
  Rev. Lett.} \textbf{\bibinfo{volume}{68}}, \bibinfo{pages}{1383}
  (\bibinfo{year}{1992}),
  \urlprefix\url{http://link.aps.org/doi/10.1103/PhysRevLett.68.1383}.

\bibitem[{\citenamefont{Suen et~al.}(1992{\natexlab{a}})\citenamefont{Suen,
  Santos, and Shayegan}}]{Suen92b}
\bibinfo{author}{\bibfnamefont{Y.~W.} \bibnamefont{Suen}},
  \bibinfo{author}{\bibfnamefont{M.~B.} \bibnamefont{Santos}},
  \bibnamefont{and} \bibinfo{author}{\bibfnamefont{M.}~\bibnamefont{Shayegan}},
  \bibinfo{journal}{Phys. Rev. Lett.} \textbf{\bibinfo{volume}{69}},
  \bibinfo{pages}{3551} (\bibinfo{year}{1992}{\natexlab{a}}),
  \urlprefix\url{https://link.aps.org/doi/10.1103/PhysRevLett.69.3551}.

\bibitem[{\citenamefont{Suen et~al.}(1992{\natexlab{b}})\citenamefont{Suen,
  Engel, Santos, Shayegan, and Tsui}}]{Suen92}
\bibinfo{author}{\bibfnamefont{Y.~W.} \bibnamefont{Suen}},
  \bibinfo{author}{\bibfnamefont{L.~W.} \bibnamefont{Engel}},
  \bibinfo{author}{\bibfnamefont{M.~B.} \bibnamefont{Santos}},
  \bibinfo{author}{\bibfnamefont{M.}~\bibnamefont{Shayegan}}, \bibnamefont{and}
  \bibinfo{author}{\bibfnamefont{D.~C.} \bibnamefont{Tsui}},
  \bibinfo{journal}{Phys. Rev. Lett.} \textbf{\bibinfo{volume}{68}},
  \bibinfo{pages}{1379} (\bibinfo{year}{1992}{\natexlab{b}}),
  \urlprefix\url{http://link.aps.org/doi/10.1103/PhysRevLett.68.1379}.

\bibitem[{\citenamefont{Suen et~al.}(1994)\citenamefont{Suen, Manoharan, Ying,
  Santos, and Shayegan}}]{Suen94b}
\bibinfo{author}{\bibfnamefont{Y.~W.} \bibnamefont{Suen}},
  \bibinfo{author}{\bibfnamefont{H.~C.} \bibnamefont{Manoharan}},
  \bibinfo{author}{\bibfnamefont{X.}~\bibnamefont{Ying}},
  \bibinfo{author}{\bibfnamefont{M.~B.} \bibnamefont{Santos}},
  \bibnamefont{and} \bibinfo{author}{\bibfnamefont{M.}~\bibnamefont{Shayegan}},
  \bibinfo{journal}{Phys. Rev. Lett.} \textbf{\bibinfo{volume}{72}},
  \bibinfo{pages}{3405} (\bibinfo{year}{1994}),
  \urlprefix\url{https://link.aps.org/doi/10.1103/PhysRevLett.72.3405}.

\bibitem[{\citenamefont{Halperin}(1983)}]{Halperin83}
\bibinfo{author}{\bibfnamefont{B.~I.} \bibnamefont{Halperin}},
  \bibinfo{journal}{Helvetica Physica Acta} \textbf{\bibinfo{volume}{56}},
  \bibinfo{pages}{75} (\bibinfo{year}{1983}), ISSN \bibinfo{issn}{0018-0238}.

\bibitem[{\citenamefont{Narasimhan and Ho}(1995)}]{Narasimhan95}
\bibinfo{author}{\bibfnamefont{S.}~\bibnamefont{Narasimhan}} \bibnamefont{and}
  \bibinfo{author}{\bibfnamefont{T.-L.} \bibnamefont{Ho}},
  \bibinfo{journal}{Phys. Rev. B} \textbf{\bibinfo{volume}{52}},
  \bibinfo{pages}{12291} (\bibinfo{year}{1995}),
  \urlprefix\url{https://link.aps.org/doi/10.1103/PhysRevB.52.12291}.

\bibitem[{\citenamefont{Shayegan et~al.}(2006)\citenamefont{Shayegan,
  De~Poortere, Gunawan, Shkolnikov, Tutuc, and Vakili}}]{Shayegan06}
\bibinfo{author}{\bibfnamefont{M.}~\bibnamefont{Shayegan}},
  \bibinfo{author}{\bibfnamefont{E.~P.} \bibnamefont{De~Poortere}},
  \bibinfo{author}{\bibfnamefont{O.}~\bibnamefont{Gunawan}},
  \bibinfo{author}{\bibfnamefont{Y.~P.} \bibnamefont{Shkolnikov}},
  \bibinfo{author}{\bibfnamefont{E.}~\bibnamefont{Tutuc}}, \bibnamefont{and}
  \bibinfo{author}{\bibfnamefont{K.}~\bibnamefont{Vakili}},
  \bibinfo{journal}{physica status solidi (b)} \textbf{\bibinfo{volume}{243}},
  \bibinfo{pages}{3629} (\bibinfo{year}{2006}), ISSN \bibinfo{issn}{1521-3951},
  \urlprefix\url{http://dx.doi.org/10.1002/pssb.200642212}.

\bibitem[{\citenamefont{Du et~al.}(2009)\citenamefont{Du, Skachko, Duerr,
  Luican, and Andrei}}]{Xu09}
\bibinfo{author}{\bibfnamefont{X.}~\bibnamefont{Du}},
  \bibinfo{author}{\bibfnamefont{I.}~\bibnamefont{Skachko}},
  \bibinfo{author}{\bibfnamefont{F.}~\bibnamefont{Duerr}},
  \bibinfo{author}{\bibfnamefont{A.}~\bibnamefont{Luican}}, \bibnamefont{and}
  \bibinfo{author}{\bibfnamefont{E.~Y.} \bibnamefont{Andrei}},
  \bibinfo{journal}{Nature} \textbf{\bibinfo{volume}{462}},
  \bibinfo{pages}{192} (\bibinfo{year}{2009}).

\bibitem[{\citenamefont{Apalkov and Chakraborty}(2010)}]{Apalkov10}
\bibinfo{author}{\bibfnamefont{V.~M.} \bibnamefont{Apalkov}} \bibnamefont{and}
  \bibinfo{author}{\bibfnamefont{T.}~\bibnamefont{Chakraborty}},
  \bibinfo{journal}{Phys. Rev. Lett.} \textbf{\bibinfo{volume}{105}},
  \bibinfo{pages}{036801} (\bibinfo{year}{2010}),
  \urlprefix\url{http://link.aps.org/doi/10.1103/PhysRevLett.105.036801}.

\bibitem[{\citenamefont{Apalkov and Chakraborty}(2011)}]{Apalkov11}
\bibinfo{author}{\bibfnamefont{V.~M.} \bibnamefont{Apalkov}} \bibnamefont{and}
  \bibinfo{author}{\bibfnamefont{T.}~\bibnamefont{Chakraborty}},
  \bibinfo{journal}{Phys. Rev. Lett.} \textbf{\bibinfo{volume}{107}},
  \bibinfo{pages}{186803} (\bibinfo{year}{2011}),
  \urlprefix\url{http://link.aps.org/doi/10.1103/PhysRevLett.107.186803}.

\bibitem[{\citenamefont{Thiebaut et~al.}(2015)\citenamefont{Thiebaut, Regnault,
  and Goerbig}}]{Thiebaut15}
\bibinfo{author}{\bibfnamefont{N.}~\bibnamefont{Thiebaut}},
  \bibinfo{author}{\bibfnamefont{N.}~\bibnamefont{Regnault}}, \bibnamefont{and}
  \bibinfo{author}{\bibfnamefont{M.~O.} \bibnamefont{Goerbig}},
  \bibinfo{journal}{Phys. Rev. B} \textbf{\bibinfo{volume}{92}},
  \bibinfo{pages}{245401} (\bibinfo{year}{2015}),
  \urlprefix\url{https://link.aps.org/doi/10.1103/PhysRevB.92.245401}.

\bibitem[{\citenamefont{Manoharan et~al.}(1996)\citenamefont{Manoharan, Suen,
  Santos, and Shayegan}}]{Manoharan96}
\bibinfo{author}{\bibfnamefont{H.~C.} \bibnamefont{Manoharan}},
  \bibinfo{author}{\bibfnamefont{Y.~W.} \bibnamefont{Suen}},
  \bibinfo{author}{\bibfnamefont{M.~B.} \bibnamefont{Santos}},
  \bibnamefont{and} \bibinfo{author}{\bibfnamefont{M.}~\bibnamefont{Shayegan}},
  \bibinfo{journal}{Phys. Rev. Lett.} \textbf{\bibinfo{volume}{77}},
  \bibinfo{pages}{1813} (\bibinfo{year}{1996}),
  \urlprefix\url{https://link.aps.org/doi/10.1103/PhysRevLett.77.1813}.

\bibitem[{\citenamefont{Shabani et~al.}(2013)\citenamefont{Shabani, Liu,
  Shayegan, Pfeiffer, West, and Baldwin}}]{Shabani13}
\bibinfo{author}{\bibfnamefont{J.}~\bibnamefont{Shabani}},
  \bibinfo{author}{\bibfnamefont{Y.}~\bibnamefont{Liu}},
  \bibinfo{author}{\bibfnamefont{M.}~\bibnamefont{Shayegan}},
  \bibinfo{author}{\bibfnamefont{L.~N.} \bibnamefont{Pfeiffer}},
  \bibinfo{author}{\bibfnamefont{K.~W.} \bibnamefont{West}}, \bibnamefont{and}
  \bibinfo{author}{\bibfnamefont{K.~W.} \bibnamefont{Baldwin}},
  \bibinfo{journal}{Phys. Rev. B} \textbf{\bibinfo{volume}{88}},
  \bibinfo{pages}{245413} (\bibinfo{year}{2013}),
  \urlprefix\url{https://link.aps.org/doi/10.1103/PhysRevB.88.245413}.

\bibitem[{\citenamefont{Hatke et~al.}(2015)\citenamefont{Hatke, Liu, Engel,
  Shyegan, Pfeiffer, West, and Baldwin}}]{Hatke15}
\bibinfo{author}{\bibfnamefont{A.~T.} \bibnamefont{Hatke}},
  \bibinfo{author}{\bibfnamefont{Y.}~\bibnamefont{Liu}},
  \bibinfo{author}{\bibfnamefont{L.~W.} \bibnamefont{Engel}},
  \bibinfo{author}{\bibfnamefont{M.}~\bibnamefont{Shyegan}},
  \bibinfo{author}{\bibfnamefont{L.~N.} \bibnamefont{Pfeiffer}},
  \bibinfo{author}{\bibfnamefont{K.~W.} \bibnamefont{West}}, \bibnamefont{and}
  \bibinfo{author}{\bibfnamefont{K.~W.} \bibnamefont{Baldwin}},
  \bibinfo{journal}{Nature Communications} \textbf{\bibinfo{volume}{6}},
  \bibinfo{pages}{7071} (\bibinfo{year}{2015}).

\bibitem[{\citenamefont{Hatke et~al.}(2014)\citenamefont{Hatke, Liu, Magill,
  Moon, Engel, Shayegan, Pfeiffer, West, and Baldwin}}]{Hatke14}
\bibinfo{author}{\bibfnamefont{A.~T.} \bibnamefont{Hatke}},
  \bibinfo{author}{\bibfnamefont{Y.}~\bibnamefont{Liu}},
  \bibinfo{author}{\bibfnamefont{B.~A.}~\bibnamefont{Magill}},
  \bibinfo{author}{\bibfnamefont{B.~H.}~\bibnamefont{Moon}},
  \bibinfo{author}{\bibfnamefont{L.~W.} \bibnamefont{Engel}},
  \bibinfo{author}{\bibfnamefont{M.}~\bibnamefont{Shyegan}},
  \bibinfo{author}{\bibfnamefont{L.~N.} \bibnamefont{Pfeiffer}},
  \bibinfo{author}{\bibfnamefont{K.~W.} \bibnamefont{West}}, \bibnamefont{and}
  \bibinfo{author}{\bibfnamefont{K.~W.} \bibnamefont{Baldwin}},
  \bibinfo{journal}{Nature Communications} \textbf{\bibinfo{volume}{5}},
  \bibinfo{pages}{4154} (\bibinfo{year}{2014}).

\bibitem[{\citenamefont{Wang et~al.}(2007)\citenamefont{Wang, Chen, Engel,
  Tsui, Tutuc, and Shayegan}}]{Wang17b}
\bibinfo{author}{\bibfnamefont{Z.}~\bibnamefont{Wang}},
  \bibinfo{author}{\bibfnamefont{Y.~P.} \bibnamefont{Chen}},
  \bibinfo{author}{\bibfnamefont{L.~W.} \bibnamefont{Engel}},
  \bibinfo{author}{\bibfnamefont{D.~C.} \bibnamefont{Tsui}},
  \bibinfo{author}{\bibfnamefont{E.}~\bibnamefont{Tutuc}}, \bibnamefont{and}
  \bibinfo{author}{\bibfnamefont{M.}~\bibnamefont{Shayegan}},
  \bibinfo{journal}{Phys. Rev. Lett.} \textbf{\bibinfo{volume}{99}},
  \bibinfo{pages}{136804} (\bibinfo{year}{2007}),
  \urlprefix\url{https://link.aps.org/doi/10.1103/PhysRevLett.99.136804}.

\bibitem[{\citenamefont{Jain}(1989)}]{Jain89}
\bibinfo{author}{\bibfnamefont{J.~K.} \bibnamefont{Jain}},
  \bibinfo{journal}{Phys. Rev. Lett.} \textbf{\bibinfo{volume}{63}},
  \bibinfo{pages}{199} (\bibinfo{year}{1989}),
  \urlprefix\url{http://link.aps.org/doi/10.1103/PhysRevLett.63.199}.

\bibitem[{\citenamefont{Jain}(2007)}]{Jain07}
\bibinfo{author}{\bibfnamefont{J.~K.} \bibnamefont{Jain}},
  \emph{\bibinfo{title}{Composite Fermions}} (\bibinfo{publisher}{Cambridge
  University Press, New York, US}, \bibinfo{year}{2007}).

\bibitem[{\citenamefont{Lopez and Fradkin}(1991)}]{Lopez91}
\bibinfo{author}{\bibfnamefont{A.}~\bibnamefont{Lopez}} \bibnamefont{and}
  \bibinfo{author}{\bibfnamefont{E.}~\bibnamefont{Fradkin}},
  \bibinfo{journal}{Phys. Rev. B} \textbf{\bibinfo{volume}{44}},
  \bibinfo{pages}{5246} (\bibinfo{year}{1991}),
  \urlprefix\url{http://link.aps.org/doi/10.1103/PhysRevB.44.5246}.

\bibitem[{\citenamefont{Halperin et~al.}(1993)\citenamefont{Halperin, Lee, and
  Read}}]{Halperin93}
\bibinfo{author}{\bibfnamefont{B.~I.} \bibnamefont{Halperin}},
  \bibinfo{author}{\bibfnamefont{P.~A.} \bibnamefont{Lee}}, \bibnamefont{and}
  \bibinfo{author}{\bibfnamefont{N.}~\bibnamefont{Read}},
  \bibinfo{journal}{Phys. Rev. B} \textbf{\bibinfo{volume}{47}},
  \bibinfo{pages}{7312} (\bibinfo{year}{1993}),
  \urlprefix\url{http://link.aps.org/doi/10.1103/PhysRevB.47.7312}.

\bibitem[{\citenamefont{Jain and Kamilla}(1997)}]{Jain97}
\bibinfo{author}{\bibfnamefont{J.~K.} \bibnamefont{Jain}} \bibnamefont{and}
  \bibinfo{author}{\bibfnamefont{R.~K.} \bibnamefont{Kamilla}},
  \bibinfo{journal}{Int. J. Mod. Phys. B} \textbf{\bibinfo{volume}{11}},
  \bibinfo{pages}{2621} (\bibinfo{year}{1997}).

\bibitem[{\citenamefont{Wu et~al.}(1993)\citenamefont{Wu, Dev, and
  Jain}}]{Wu93}
\bibinfo{author}{\bibfnamefont{X.~G.} \bibnamefont{Wu}},
  \bibinfo{author}{\bibfnamefont{G.}~\bibnamefont{Dev}}, \bibnamefont{and}
  \bibinfo{author}{\bibfnamefont{J.~K.} \bibnamefont{Jain}},
  \bibinfo{journal}{Phys. Rev. Lett.} \textbf{\bibinfo{volume}{71}},
  \bibinfo{pages}{153} (\bibinfo{year}{1993}),
  \urlprefix\url{http://link.aps.org/doi/10.1103/PhysRevLett.71.153}.

\bibitem[{\citenamefont{Park and Jain}(1998)}]{Park98}
\bibinfo{author}{\bibfnamefont{K.}~\bibnamefont{Park}} \bibnamefont{and}
  \bibinfo{author}{\bibfnamefont{J.~K.} \bibnamefont{Jain}},
  \bibinfo{journal}{Phys. Rev. Lett.} \textbf{\bibinfo{volume}{80}},
  \bibinfo{pages}{4237} (\bibinfo{year}{1998}),
  \urlprefix\url{http://link.aps.org/doi/10.1103/PhysRevLett.80.4237}.

\bibitem[{\citenamefont{Mandal and Ravishankar}(1996)}]{Mandal96}
\bibinfo{author}{\bibfnamefont{S.~S.} \bibnamefont{Mandal}} \bibnamefont{and}
  \bibinfo{author}{\bibfnamefont{V.}~\bibnamefont{Ravishankar}},
  \bibinfo{journal}{Phys. Rev. B} \textbf{\bibinfo{volume}{54}},
  \bibinfo{pages}{8699} (\bibinfo{year}{1996}),
  \urlprefix\url{http://link.aps.org/doi/10.1103/PhysRevB.54.8699}.

\bibitem[{\citenamefont{Girvin and MacDonald}(2007)}]{Girvin07}
\bibinfo{author}{\bibfnamefont{S.~M.} \bibnamefont{Girvin}} \bibnamefont{and}
  \bibinfo{author}{\bibfnamefont{A.~H.} \bibnamefont{MacDonald}},
  \emph{\bibinfo{title}{Multicomponent Quantum Hall Systems: The Sum of Their
  Parts and More}} (\bibinfo{publisher}{Wiley-VCH Verlag GmbH},
  \bibinfo{year}{2007}), pp. \bibinfo{pages}{161--224}, ISBN
  \bibinfo{isbn}{9783527617258},
  \urlprefix\url{http://dx.doi.org/10.1002/9783527617258.ch5}.

\bibitem[{\citenamefont{Haldane}(1983)}]{Haldane83}
\bibinfo{author}{\bibfnamefont{F.~D.~M.} \bibnamefont{Haldane}},
  \bibinfo{journal}{Phys. Rev. Lett.} \textbf{\bibinfo{volume}{51}},
  \bibinfo{pages}{605} (\bibinfo{year}{1983}),
  \urlprefix\url{http://link.aps.org/doi/10.1103/PhysRevLett.51.605}.

\bibitem[{\citenamefont{P\'erez-Garrido and Moore}(1999)}]{thomsonproblem}
\bibinfo{author}{\bibfnamefont{A.}~\bibnamefont{P\'erez-Garrido}}
  \bibnamefont{and} \bibinfo{author}{\bibfnamefont{M.~A.} \bibnamefont{Moore}},
  \bibinfo{journal}{Phys. Rev. B} \textbf{\bibinfo{volume}{60}},
  \bibinfo{pages}{15628} (\bibinfo{year}{1999}),
  \urlprefix\url{https://link.aps.org/doi/10.1103/PhysRevB.60.15628}.

\bibitem[{\citenamefont{Erber and Hockney}(1995)}]{eqmconfigsErber}
\bibinfo{author}{\bibfnamefont{T.}~\bibnamefont{Erber}} \bibnamefont{and}
  \bibinfo{author}{\bibfnamefont{G.~M.} \bibnamefont{Hockney}},
  \bibinfo{journal}{Phys. Rev. Lett.} \textbf{\bibinfo{volume}{74}},
  \bibinfo{pages}{495} (\bibinfo{year}{1995}).

\bibitem[{\citenamefont{Tarnai}(2002)}]{tammesproblem}
\bibinfo{author}{\bibfnamefont{T.}~\bibnamefont{Tarnai}}, \bibinfo{journal}{J.
  Struct. Chem.} \textbf{\bibinfo{volume}{13}}, \bibinfo{pages}{289}
  (\bibinfo{year}{2002}), ISSN \bibinfo{issn}{1572-9001},
  \urlprefix\url{http://dx.doi.org/10.1023/A:1015859822819}.

\bibitem[{\citenamefont{Bowick et~al.}()\citenamefont{Bowick, Cecka, C.,
  Middleton, and Zielnicki}}]{Bowick}
\bibinfo{author}{\bibfnamefont{M.}~\bibnamefont{Bowick}},
  \bibinfo{author}{\bibnamefont{Cecka}},
  \bibinfo{author}{\bibfnamefont{L.}~\bibnamefont{C.}, \bibfnamefont{Gioma}},
  \bibinfo{author}{\bibfnamefont{A.}~\bibnamefont{Middleton}},
  \bibnamefont{and}
  \bibinfo{author}{\bibfnamefont{K.}~\bibnamefont{Zielnicki}},
  \emph{\bibinfo{title}{Thomson problem}},
  \bibinfo{howpublished}{\url{http://thomson.phy.syr.edu/}}, \bibinfo{note}{accessed: 2017-08-17}.

\bibitem[{\citenamefont{Wales et~al.}(2009)\citenamefont{Wales, McKay, and
  Altschuler}}]{defectmotifs}
\bibinfo{author}{\bibfnamefont{D.~J.} \bibnamefont{Wales}},
  \bibinfo{author}{\bibfnamefont{H.}~\bibnamefont{McKay}}, \bibnamefont{and}
  \bibinfo{author}{\bibfnamefont{E.~L.} \bibnamefont{Altschuler}},
  \bibinfo{journal}{Phys. Rev. B} \textbf{\bibinfo{volume}{79}},
  \bibinfo{pages}{224115} (\bibinfo{year}{2009}),
  \urlprefix\url{https://link.aps.org/doi/10.1103/PhysRevB.79.224115}.

\bibitem[{\citenamefont{Wales}(2004)}]{walesbook}
\bibinfo{author}{\bibfnamefont{D.}~\bibnamefont{Wales}},
  \emph{\bibinfo{title}{Energy Landscapes}} (\bibinfo{publisher}{Cambridge
  University Press}, \bibinfo{address}{University Printing House, Shaftesbury
  Road, Cambridge, CB2 8BS}, \bibinfo{year}{2004}), \bibinfo{edition}{1st} ed.,
  ISBN \bibinfo{isbn}{9780521814157}.

\bibitem[{\citenamefont{Doye et~al.}(1998)\citenamefont{Doye, Wales, and
  Miller}}]{lj38}
\bibinfo{author}{\bibfnamefont{J.~P.~K.} \bibnamefont{Doye}},
  \bibinfo{author}{\bibfnamefont{D.~J.} \bibnamefont{Wales}}, \bibnamefont{and}
  \bibinfo{author}{\bibfnamefont{M.~A.} \bibnamefont{Miller}},
  \bibinfo{journal}{J. Chem. Phys.} \textbf{\bibinfo{volume}{109}},
  \bibinfo{pages}{8143} (\bibinfo{year}{1998}),
  \eprint{http://dx.doi.org/10.1063/1.477477},
  \urlprefix\url{http://dx.doi.org/10.1063/1.477477}.

\bibitem[{\citenamefont{Majumdar and Martin}(2006)}]{nomin}
\bibinfo{author}{\bibfnamefont{S.~N.} \bibnamefont{Majumdar}} \bibnamefont{and}
  \bibinfo{author}{\bibfnamefont{O.~C.} \bibnamefont{Martin}},
  \bibinfo{journal}{Phys. Rev. E} \textbf{\bibinfo{volume}{74}},
  \bibinfo{pages}{061112} (\bibinfo{year}{2006}),
  \urlprefix\url{https://link.aps.org/doi/10.1103/PhysRevE.74.061112}.

\bibitem[{\citenamefont{Kusumaatmaja and
  Wales}(2013)}]{defectmotifsmeancurvature}
\bibinfo{author}{\bibfnamefont{H.}~\bibnamefont{Kusumaatmaja}}
  \bibnamefont{and} \bibinfo{author}{\bibfnamefont{D.~J.} \bibnamefont{Wales}},
  \bibinfo{journal}{Phys. Rev. Lett.} \textbf{\bibinfo{volume}{110}},
  \bibinfo{pages}{165502} (\bibinfo{year}{2013}),
  \urlprefix\url{https://link.aps.org/doi/10.1103/PhysRevLett.110.165502}.

\bibitem[{\citenamefont{Li and Scheraga}(1987)}]{Li6611}
\bibinfo{author}{\bibfnamefont{Z.}~\bibnamefont{Li}} \bibnamefont{and}
  \bibinfo{author}{\bibfnamefont{H.~A.} \bibnamefont{Scheraga}},
  \bibinfo{journal}{Proceedings of the National Academy of Sciences}
  \textbf{\bibinfo{volume}{84}}, \bibinfo{pages}{6611} (\bibinfo{year}{1987}),
  ISSN \bibinfo{issn}{0027-8424},
  \eprint{http://www.pnas.org/content/84/19/6611.full.pdf},
  \urlprefix\url{http://www.pnas.org/content/84/19/6611}.

\bibitem[{\citenamefont{Wales and Doye}(1997)}]{basinhopping}
\bibinfo{author}{\bibfnamefont{D.~J.} \bibnamefont{Wales}} \bibnamefont{and}
  \bibinfo{author}{\bibfnamefont{J.~P.~K.} \bibnamefont{Doye}},
  \bibinfo{journal}{J. Phys. Chem. A} \textbf{\bibinfo{volume}{101}},
  \bibinfo{pages}{5111} (\bibinfo{year}{1997}),
  \eprint{http://dx.doi.org/10.1021/jp970984n},
  \urlprefix\url{http://dx.doi.org/10.1021/jp970984n}.

\bibitem[{\citenamefont{Ceulemans and Fowler}(1995)}]{eulerstheorem}
\bibinfo{author}{\bibfnamefont{A.}~\bibnamefont{Ceulemans}} \bibnamefont{and}
  \bibinfo{author}{\bibfnamefont{P.~W.} \bibnamefont{Fowler}},
  \bibinfo{journal}{J. Chem. Soc. Faraday Trans.}
  \textbf{\bibinfo{volume}{91}}, \bibinfo{pages}{3089} (\bibinfo{year}{1995}),
  \urlprefix\url{http://dx.doi.org/10.1039/FT9959103089}.

\bibitem[{\citenamefont{Wales and
  Ulker}(2006{\natexlab{a}})}]{sphericalcrystals}
\bibinfo{author}{\bibfnamefont{D.~J.} \bibnamefont{Wales}} \bibnamefont{and}
  \bibinfo{author}{\bibfnamefont{S.}~\bibnamefont{Ulker}},
  \bibinfo{journal}{Phys. Rev. B} \textbf{\bibinfo{volume}{74}},
  \bibinfo{pages}{212101} (\bibinfo{year}{2006}{\natexlab{a}}),
  \urlprefix\url{https://link.aps.org/doi/10.1103/PhysRevB.74.212101}.

\bibitem[{\citenamefont{Wales}(2017)}]{software}
\bibinfo{author}{\bibfnamefont{D.}~\bibnamefont{Wales}},
  \emph{\bibinfo{title}{Wales group software}},
  \bibinfo{howpublished}{\url{http://www-wales.ch.cam.ac.uk/software.html}}
  (\bibinfo{year}{2017}), \bibinfo{note}{accessed: 2017-05-03}.

\bibitem[{\citenamefont{Trygubenko and Wales}(2004)}]{L-BFGS}
\bibinfo{author}{\bibfnamefont{S.~A.} \bibnamefont{Trygubenko}}
  \bibnamefont{and} \bibinfo{author}{\bibfnamefont{D.~J.} \bibnamefont{Wales}},
  \bibinfo{journal}{J. Chem. Phys.} \textbf{\bibinfo{volume}{120}},
  \bibinfo{pages}{2082} (\bibinfo{year}{2004}),
  \eprint{http://dx.doi.org/10.1063/1.1636455},
  \urlprefix\url{http://dx.doi.org/10.1063/1.1636455}.

\bibitem[{\citenamefont{Schebarchov and Wales}(2015)}]{combsearch}
\bibinfo{author}{\bibfnamefont{D.}~\bibnamefont{Schebarchov}} \bibnamefont{and}
  \bibinfo{author}{\bibfnamefont{D.~J.} \bibnamefont{Wales}},
  \bibinfo{journal}{Phys. Chem. Chem. Phys.} \textbf{\bibinfo{volume}{17}},
  \bibinfo{pages}{28331} (\bibinfo{year}{2015}),
  \urlprefix\url{http://dx.doi.org/10.1039/C5CP01198A}.

\bibitem[{\citenamefont{Wales and Scheraga}(1999)}]{globopt}
\bibinfo{author}{\bibfnamefont{D.~J.} \bibnamefont{Wales}} \bibnamefont{and}
  \bibinfo{author}{\bibfnamefont{H.~A.} \bibnamefont{Scheraga}},
  \bibinfo{journal}{Science} \textbf{\bibinfo{volume}{285}},
  \bibinfo{pages}{1368} (\bibinfo{year}{1999}), ISSN \bibinfo{issn}{0036-8075},
  \eprint{http://science.sciencemag.org/content/285/5432/1368.full.pdf},
  \urlprefix\url{http://science.sciencemag.org/content/285/5432/1368}.

\bibitem[{\citenamefont{Park et~al.}(1999)\citenamefont{Park, Meskini, and
  Jain}}]{Park99b}
\bibinfo{author}{\bibfnamefont{K.}~\bibnamefont{Park}},
  \bibinfo{author}{\bibfnamefont{N.}~\bibnamefont{Meskini}}, \bibnamefont{and}
  \bibinfo{author}{\bibfnamefont{J.}~\bibnamefont{Jain}}, \bibinfo{journal}{J.
  Phys. Condens. Mat.} \textbf{\bibinfo{volume}{11}} (\bibinfo{year}{1999}).

\end{thebibliography}


\end{document}